\documentclass[
aps,
prc,
superscriptaddress,
floatfix,
nofootinbib,
twocolumn
]{revtex4-2}

\usepackage{amsmath,amssymb,amsfonts,physics,graphicx,float}
\graphicspath{{pic/}}
\usepackage[setpagesize=false]{hyperref}
\allowdisplaybreaks[4]

\usepackage[labelformat=simple]{subcaption} 

\usepackage{ulem}
\usepackage{color}
\definecolor{ar}{rgb}{1.0, 0.01, 0.24}
\definecolor{al}{rgb}{0.82, 0.1, 0.26}
\definecolor{ev}{rgb}{0.56, 0.0, 1.0}

\newcommand{\mir}{\mathrm{mir}}

\begin{document}

\title{
Parity doublet model for baryon octets: 
ground states saturated by good diquarks and the role of bad diquarks for excited states 
}
\author{Bikai Gao}
\email{gaobikai@hken.phys.nagoya-u.ac.jp}
\affiliation{Department of Physics, Nagoya University, Nagoya 464-8602, Japan}

\author{Toru Kojo}
\email{torujj@nucl.phys.tohoku.ac.jp}
\affiliation{Department of Physics, Tohoku University, Sendai 980-8578, Japan}

\author{Masayasu Harada}
\email{harada@hken.phys.nagoya-u.ac.jp}
\affiliation{Kobayashi-Maskawa Institute for the Origin of Particles and the Universe, Nagoya University, Nagoya, 464-8602, Japan}
\affiliation{Department of Physics, Nagoya University, Nagoya 464-8602, Japan}
\affiliation{Advanced Science Research Center, Japan Atomic Energy Agency, Tokai 319-1195, Japan}

\date{\today}

\begin{abstract}
Parity doublet model is an effective chiral model that includes the chiral variant and invariant masses of baryons.
The chiral invariant mass has large impacts on the density dependence of models which can be constrained by neutron star observations. 
In the previous work, models of two-flavors  have been considered up to a few times nuclear saturation density, 
but in such dense region it is also necessary to consider hyperons.
With the chiral invariant masses 
baryons can stay massive in extreme environments (e.g., neutron stars) where the chiral symmetry restoration takes place.
In this work, we generalize the previous $\mbox{SU(2)}_L \times \mbox{SU(2)}_R$ parity models of nucleons 
to $\mbox{SU(3)}_L \times \mbox{SU(3)}_R$ models of the baryon octet, within the linear realization of the chiral symmetry. 
The major problem in constructing such models has been too many candidates for the chiral representations of baryons.
Motivated by the concepts of diquarks and the mended symmetry,
we choose the $(3_L, \bar{3}_R) + (\bar{3}_L, 3_R)$, $(3_L, 6_R) + (6_L, 3_R)$ and $(1_L, 8_R) + (8_L, 1_R)$ representations
and use quark diagrams to constrain the possible types of Yukawa interactions. 
The masses of the baryon octets
for positive and negative baryons up to the first excitations are successfully reproduced. 
As expected from the diquark considerations,
the ground state baryons are well dominated by $(3_L, \bar{3}_R) + (\bar{3}_L, 3_R)$ and $(1_L, 8_R) + (8_L, 1_R)$ representations,
while the excited states require $(3_L, 6_R) + (6_L, 3_R)$ representations. Important applications of our model are the chiral restoration for strange quarks at large density
and the continuity of diquarks from hadronic to quark matter.
We also address the problem of large Yukawa couplings which are enhanced in three-flavor construction.
\end{abstract}

\maketitle


\section{Introduction}


Understanding of the origin of the nucleon mass is one of the most fundamental subjects in hadron physics. Traditionally, the chiral symmetry 
 in quantum chromodynamics (QCD) 
and its spontaneous symmetry breaking (SSB) \cite{Weinberg:1966kf,Weinberg:1969hw,Weinberg:1990xn,Weinberg:2010bq} are often used to study the nucleon mass.
The order parameter of the SSB is the chiral condensate $\langle \bar{q} q\rangle$, which is made of quark-antiquark pairs with different chirality \cite{Nambu:1961tp,Nambu:1961fr,Hatsuda:1994pi}. 
The existence of chiral condensate breaks the chiral symmetry because it couples quarks of different chirality and in vacuum it has a nonzero value. 
In certain extreme conditions, such as high-density or high-temperature region, the chiral condensate may vanish with the restoration of chiral symmetry
in the thermodynamic state.

The linear sigma model (LSM) is an effective model broadly used for investigating the SSB. In the LSM, the order parameter of SSB is the expectation values of the scalar field $\langle \sigma \rangle \propto \langle \bar{q} q \rangle$. In the traditional hadronic model with LSM, the nucleon mass $m_{N}$ is considered to be generated mainly by chiral condensates. An  interesting consequence of this perspective is that the nucleon mass vanishes in the high density or temperature region where the chiral symmetry should be restored.

However, the traditional view of nucleon masses being completely originated from the chiral condensate is being reconsidered, 
due to the insights gained from lattice QCD simulations
\cite{Aarts:2015mma,Aarts:2017rrl, Aarts:2017iai, Aarts:2018glk, Aarts:2019dot}.  
These lattice QCD results have indicated the possibility of a chiral-invariant mass, denoted as $m_0$, 
which seems to exist apart from the conventional chiral-variant mass which has  dependence over  the density or temperature. 
This novel concept of a chiral-invariant mass suggests the nucleon mass would keep a finite value even when the chiral symmetry is restored. 
The concept of the chiral invariant mass is naturally incorporated 
in a parity doublet model (PDM) for nucleons in which the ordinary nucleon and its parity partner form a doublet structure \cite{Detar:1988kn,Jido:1998av,Jido:1999hd,Jido:2001nt,Nagata:2008xf,Gallas:2009qp,Gallas:2013ipa,Motohiro:2015taa}. 
Within this framework, nucleon masses in the PDM are found to be less sensitive to the chiral condensate than in  the conventional LSM. 
This changing of the sensitivity has far-reaching consequences, especially when we construct the equation of state (EOS) for nuclear and neutron star (NS) matter. 
The introduction of chiral invariant mass leads to significant changes in the coupling constants which, in turn, greatly affect the stiffness of the EOS. 

In the context of applications to NS phenomenology, the nuclear EOS within the PDM is often extended to densities beyond the nuclear saturation density $n_{0}$ \cite{Sasaki:2010bp,Eser:2023oii}. 
This extrapolation has been achieved in various ways. 
One approach involves a straightforward extrapolation of the PDM EOS beyond $n_0 \simeq 0.16\, {\rm fm}^{-3}$,
 as studied in Ref.~\cite{Yamazaki:2019tuo}. 
Another method combines the PDM EOS with a quark model, assuming a quark-hadron crossover, 
which allows for a smooth transition from hadronic matter to quark matter \cite{Baym:2017whm,Minamikawa:2020jfj,Minamikawa:2021fln,Gao:2022klm,Minamikawa:2023eky,Marczenko:2019trv}. 
This hybrid approach, where the PDM EOS is employed up to densities around 2-3$n_0$ and interpolate with the quark EOS at $\geq 5 n_0$ via polynomial interpolants to obtain the unified EOS. This unified EOS is valuable for modeling the behavior of dense matter within neutron stars and can provide us important insights into the intermediate density region. However, the validity of a purely hadronic picture at densities above 2$n_0$ is questionable because hyperons, strange baryons containing strange quarks, may enter the matter and affect its properties. Taking this possibility into consideration, it becomes necessary to explore the effect of strange quark at finite density region. 
In previous study for the nuclear matter domain, we included the strange quark effects through the Kobayashi-Maskawa-'t Hooft (KMT) interactions and constructed a three-flavor mesonic Lagrangian made of scalar and vector mesons \cite{Gao:2022klm}. The corresponding unified EOS was confronted with NS constraints from the NS merger GW170817 \cite{PhysRevLett.119.161101,LIGOScientific:2017ync,LIGOScientific:2018cki}, the millisecond pulsar PSR J0030+0451 \cite{Riley:2021pdl}, and the maximum mass constraint from the millisecond pulsar PSR J0740+6620 \cite{Miller:2021qha}. Based on the above NS observation data, we constrained $m_0$ to the range $400\, \mathrm{MeV} \lesssim m_0 \lesssim 700\, \mathrm{MeV} $ which is more relaxed than the one obtained in Ref.~\cite{Minamikawa:2020jfj}.

In the analyses in Refs.~\cite{Minamikawa:2020jfj,Minamikawa:2021fln,Gao:2022klm,Minamikawa:2023eky}, we have assumed that the hyperons do not appear in the density region below $2n_0$. 
On the other hand, the analysis in Ref.~\cite{Minamikawa:2021fln} shows that strangeness number density starts to appear in the crossover region at $n_B \gtrsim 2 n_0$.
This indicates that the hyperons might appear even below $2n_0$.
To clarify this, we need to construct a parity doublet model including hyperons.
There were several analyses of chiral models with the parity doublet structure based on the SU(3)$_L \otimes$ SU(3)$_R$ chiral symmetry~\cite{Chen:2009sf,Chen:2010ba,Chen:2011rh,Nishihara:2015fka,Dmitrasinovic:2016hup,Minamikawa:2023ypn}.
Since the SU(3) flavor octet baryons appear from the  chiral representation of $(3_L, \bar{3}_R)$, $(8_{L}, 1_{R})$, $(3_{L}, 6_{R})$ with $L \leftrightarrow R$, all of three chiral representations are included in most of these analyses.

In the previous analysis \cite{Minamikawa:2023ypn},
we adopted the dynamical assumption and included only $(3_L, \bar{3}_R)$ and $(8_{L}, 1_{R})$ (and $L \leftrightarrow R$) representations which are made from so called ``good diquark'' without including $(3_{L}, 6_{R})$ (and $L \leftrightarrow R$) from the ``bad diquark''.
Based on the quark diagrams, we constructed Yukawa interactions of baryons to the chiral field $\Sigma$, the vacuum expectation value (VEV)
 of which spontaneously breaks the chiral symmetry.
Then, we have proven that the first order interactions cannot reproduce the correct mass ordering of octet baryons in the ground state, $m_N < m_\Lambda \sim m_\Sigma < m_\Xi$.
We also showed that the inclusion of the second order Yukawa interactions can generate the correct mass ordering of the ground state baryons.
However, in the excited states, the mass ordering appears to be incorrect for the wide domain of model parameters which we have exhaustively explored.
This indicates that models made only of baryons with good diquarks, even after including the Yukawa interactions up to the second order,
misses some qualitative features of the baryon spectra.

In the present analysis, we include the $(3_L, \bar{3}_R)$, $(8_{L}, 1_{R})$, $(3_{L}, 6_{R})$ (and $L\leftrightarrow R$) chiral representations
and construct the Yukawa interactions based on the quark diagram as in Ref.~\cite{Minamikawa:2023ypn}.
We will show that the mass ordering of the excited states is correctly reproduced within the reasonable range of parameters; the problem found in Ref.~\cite{Minamikawa:2023ypn} is solved.
Furthermore, we study the composition of chiral representations in the ground and excited baryons.
We found that the ground states are 
dominated by the representations of $(3_L, \bar{3}_R)$ and $(8_{L}, 1_{R})$ made of good diquarks.
This finding is consistent with the conventional arguments in the hadron spectroscopy.
Meanwhile the excited baryons contain the substantial component of $(3_{L}, 6_{R})$ with bad diquarks.
This finding is rather new; 
discussions on bad diquarks in our paper should not be confused with the mass splitting between the baryon octet and decuplet (e.g., the $N$-$\Delta$ splitting)
for which diquarks offer simple explanations.
Instead, our work addresses diquarks for the mass splitting between the ground and excited states {\it within} the baryon octet;
here the utility of diquarks has not been established.

This paper is structured as follows.
In Sec.\ref{Model_construction}, the chiral representations of 
$(3_L,\bar{3}_R)+(\bar{3}_L,3_R)$, $(3_L, 6_R) + (6_L, 3_R)$ and $(8_L,1_R)+(1_L,8_R)$ 
for octet baryons are defined. 
In Sec.\ref{Yukawa}, we  construct an effective Lagrangian for baryons including first order   Yukawa interactions.
In Sec.\ref{Numerical}, we perform numerical fit of baryon spectra.
Sec.\ref{sec-summary} is devoted to the summary. 





\section{Model construction}
\label{Model_construction}

%
\begin{figure*}[t]\centering
\begin{subfigure}{0.3\hsize}\centering
	\includegraphics[width=0.8\hsize]{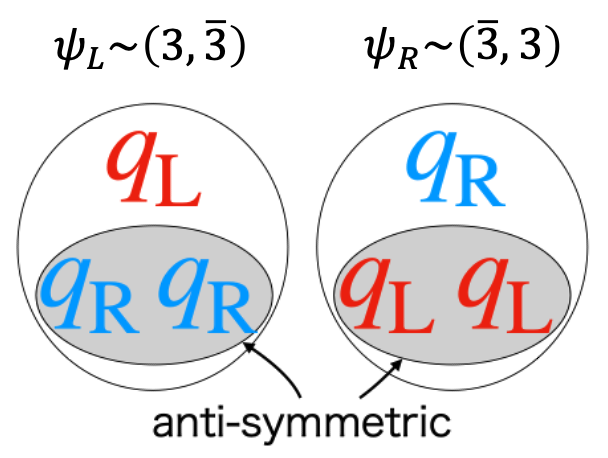}
	\caption{}
	\label{fig-baryon-psi}
\end{subfigure}
\begin{subfigure}{0.3\hsize}\centering
	\includegraphics[width=0.8\hsize]{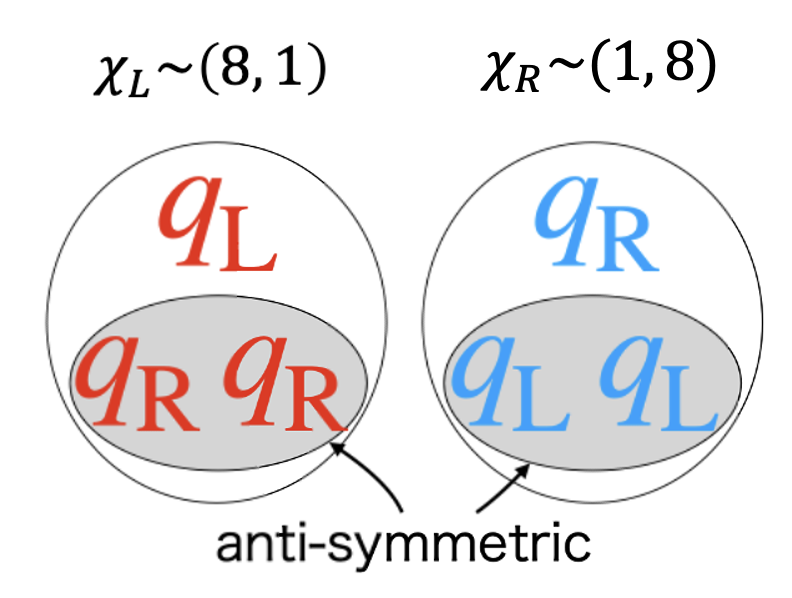}
	\caption{}
	\label{fig-baryon-chi}
\end{subfigure}
\begin{subfigure}{0.3\hsize}\centering
	\includegraphics[width=0.8\hsize]{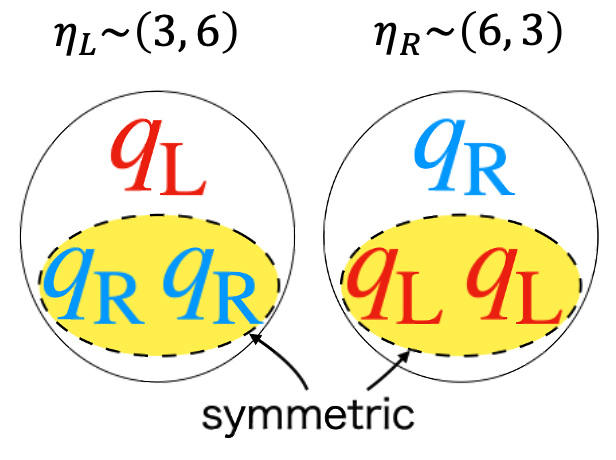}
	\caption{}
	\label{fig-baryon-eta}
\end{subfigure}
\caption[]{
Quark contents for three  
baryon representations: 
(a) $(3_L,\bar3_R)+(\bar{3}_L,3_R)$, 
(b) $(8_L,1_R)+(1_L,8_R)$
and
(c) $(3_L,6_R)+(6_L,3_R)$. 
The gray shaded diquark indicates flavor antisymmetric representation, 
while the yellow indicates symmetric one. 
}
\label{fig-baryon}
\end{figure*}
%

In the chiral SU(3)$_{L} \times$ SU(3)$_{R}$ symmetry, there are so many chiral representations for baryons. 
In order to keep only minimal set of necessary fields,
we first consider the quark substructure for each representation
and then choose baryon fields which are supposed to be relevant from low to intermediate energies.
We further include the parity doublet structure to consider the chiral invariant mass.

The chiral representations of quarks under  chiral  SU(3)$_{L} \times$ SU(3)$_{R}$  symmetry  are written as
\begin{align}
(q_{L})^{l} &\sim (3_{L}, 1_{R})  \sim (u_{L}, d_{L}, s_{L})^{l}, \\
(q_{R})^{r} &\sim (1_{L}, 3_{R})  \sim (u_{R}, d_{R}, s_{R})^{r},
\end{align}
where the  indices $l = 1,2,3$ and $r=1,2,3$ are for SU(3)$_L$ and SU(3)$_R$ symmetries, respectively.   
Since a baryon can be expressed as a direct product of three quarks, the possible combinations are $(q_L)^3, (q_L)^2 q_R, q_L (q_R)^2$, and $q_R^3$.
It is sufficient to consider
the following possibilities for the chiral representations of baryons:
\begin{equation}
\begin{gathered}
q_{\mathrm{L}} \times \left(q_{\mathrm{L}} \times q_{\mathrm{L}} + q_{\mathrm{R}} \times q_{\mathrm{R}}\right) \sim \\
(1_L,1_R)+(8_L,1_R)+(8_L,1_R) \\
+(10_L,1_R)+(3_L, \overline{3}_R)+(3_L,6_R)\ ,
\end{gathered}
\end{equation}
and $L \leftrightarrow R$. 
The octet baryons are included in 
$(3_L, \bar{3}_R)$, $(8_L, 1_R)$, and $(3_L, 6_R)$ 
which are illustrated in Fig.~\ref{fig-baryon}. 
Following Ref.~\cite{Nishihara:2015fka}, in the framework of the parity doublet model
we introduce the corresponding baryon  fields 
$\psi$, $\chi$, $\eta$ and its parity doubling partners $\psi^{{\rm mir}}$, $\chi^{{\rm mir}}$, $\eta^{{\rm mir}}$ as
\begin{align}
\psi_{L} \sim (3_L, \bar{3}_R)&, \quad \psi_{L}^{{\rm mir}} \sim (\bar{3}_L, 3_R)\, \\
\chi_{L} \sim (8_L, 1_R) &, \quad \chi_{L}^{{\rm mir}} \sim (1_L, 8_R) \,,\\
\eta_{L} \sim (3_L, 6_R) &, \quad \eta_{L}^{{\rm mir}} \sim (6_L, 3_R) \,.
\end{align} 
Here the indices $L$ and $R$ in the subscript of $\psi, \chi$, and $\eta$ represents the chirality in U(1)$_A$, e.g., $\gamma_5 \psi_{R,L} = \pm \psi_{R,L}$.
The right-handed fields are also defined in the similar way. Under the parity transformation ($\mathcal{P}$) and the charge conjugation ($\mathcal{C}$), 
these fields transform as
\begin{equation}
\begin{array}{ll}
\Psi_{ l,  r} \stackrel{\mathcal{P}}{\rightarrow} \gamma_0 \Psi_{ r,  l}, & \Psi^{\mir}_{ l,  r} \stackrel{\mathcal{P}}{\rightarrow}-\gamma_0 \Psi^{\mir}_{ r,  l}, \\
\Psi_{ l,  r} \stackrel{\mathcal{C}}{\rightarrow} C\left(\bar{\psi}_{ r,  l}\right)^T, & \Psi^{\mir}_{ l,  r} \stackrel{\mathcal{C}}{\rightarrow}-C\left(\bar{\Psi}^{\mir}_{ r,  l}\right)^T \ ,
\end{array}
\end{equation}
for $\Psi = \psi, \chi, \eta$ with $\mathcal{C} = i\gamma^{2}\gamma^{0}$. 
This assignment for the parity is called the mirror assignment.

The irreducible representations of chiral SU(3)$_{L} \times$ SU(3)$_{R}$,
$\psi, \chi, \eta$, are not generally irreducible representations of the flavor SU(3)$_{V=L+R}$.
Here we briefly look at how $\psi, \chi, \eta$ can be expressed in terms of the conventional flavor SU(3)$_{V}$,
as we will eventually consider the chiral SSB, SU(3)$_{L} \times$ SU(3)$_{R} \rightarrow $ SU(3)$_{V}$.

The field $\psi_L \sim (3_L, \bar{3}_R)$ can be written as
\begin{equation}
\big( \psi_L \big)^a_{ \bar{\alpha} }
= \frac{\, \delta^a_{\bar{\alpha}} \,}{\, \sqrt{3} \,} \Lambda_{0L} 
+ (B_{\psi L})^a_{\bar{\alpha}} \ , 
\end{equation}
where $a = 1, 2, 3$ are for the fundamental representations in SU(3)$_{L}$ and $\bar{\alpha}= 1, 2, 3$ for the anti-fundamental representations in the SU(3)$_{R}$.  
The $\Lambda_0$ is the flavor singlet $\Lambda$, while $B$ is the flavor octet
\begin{align}
B_\psi = \left(\begin{array}{ccc}
\frac{1}{\sqrt{2}} \Sigma^0+\frac{1}{\sqrt{6}} \Lambda & \Sigma^{+} & p \\
\Sigma^{-} & \frac{-1}{\sqrt{2}} \Sigma^0+\frac{1}{\sqrt{6}} \Lambda & n \\
\Xi^{-} & \Xi^0 & \frac{-2}{\sqrt{6}} \Lambda
\end{array}\right)_\psi \,.
\end{align}
We can repeat similar assignments for $\chi_{L} \sim (8_L, 1_R)$, 
\begin{equation}
(\chi_L )^{a}_{\bar{b}} = (B_{\chi L})^{a}_{\bar{b}} \ ,
\end{equation}
where $B_\chi$ has the same matrix structure as $B_\psi$.
The indices $\bar{b} = 1,.,3$  represents $\bar{3}_L$.
We note that $B_\psi$ and $B_\chi$ transforms differently in SU(3)$_{L} \times$ SU(3)$_{R}$,
but show the same transformation under the flavor SU(3)$_V$;
in model having only the SU(3)$_V$ symmetry, they are degenerate.
Finally the field $\eta_L \sim (3_L, 6_R)$ can be labelled as
\begin{equation}
( \eta_L )^{ [a, \alpha \beta] }
= \frac{1}{\sqrt{6}}\left(\epsilon^{\alpha a \gamma } \delta_b^\beta+\epsilon^{\beta a \gamma} \delta_b^\alpha\right) (B_{\eta L})^b_{\gamma} \\
\end{equation}
with $B_{\eta}$ having the same structure as $B_{\psi,\chi}$.
We can write the right-handed fields in the same way.

For mirror fields, we just exchange the index for the U(1)$_A$ to get the same assignment rules.
For example the field $\psi_R^{\rm mir} \sim (3_L, \bar{3}_R)$ can be written as
\begin{equation}
\big( \psi_R^{\mir} \big)^a_{ \bar{\alpha} }
= \frac{\, \delta^a_{\bar{\alpha}} \,}{\, \sqrt{3} \,} \Lambda_{0R} 
+ (B_{\psi R}^{\rm mir})^a_{\bar{\alpha}} \,.
\end{equation}
The fields $\psi_L$ and $\psi_R^{\rm mir}$ follows the same transformation under the SU(3)$_{L} \times \,$SU(3)$_{R}$, i.e.,
 $\psi_{L}\rightarrow g_{L}\psi_{L}g_{R}^{\dagger}$ and $\psi_{R}^{{\rm mir}} \rightarrow g_{L}\psi_{R}^{{\rm mir}} g_{R}^{\dagger}$.
 This allows the mass term $\tr \big[ \bar{\psi}_L \psi_R^{\rm mir} \big]$ which is invariant under SU(3)$_{L} \times$ SU(3)$_{R}$.

The introduction of the mirror fields enables us to write down the mixing terms as
\begin{align}
\mathcal{L}_{m_0} =& - m^{\psi}_{0}  {\rm tr} \left( \bar{\psi}_{R}\psi^{{\rm mir}}_{L} - \bar{\psi}_{L}\psi^{{\rm mir}}_{R} \right) + h.c \nonumber\\
&-m_{0}^{\chi}{\rm tr} \left( \bar{\chi}_{R}\chi_{L}^{{\rm mir}} - \bar{\chi}_{L}\chi_{R}^{{\rm mir}} \right) + h.c  \nonumber\\
&-m_{0}^{\eta} \tr \left( \bar{\eta}_{R}\eta_{L}^{{\rm mir}} - \bar{\eta}_{L} \eta_{R}^{{\rm mir}} \right) + h.c \,.
\end{align}
The parameter $m_{0}^{\psi, \chi, \eta}$ corresponds to the chiral invariant  masses of $\psi$, $\chi$ and $\eta$ fields, respectively.  
Note here that the $\psi$ and $\chi$ fields contain flavor-antisymmetric diquarks which are called {\it ``good" } diquarks, 
while $\eta$ field  
contains flavor-symmetric diquark called {\it ``bad"} diquarks. 
We assume that baryons in representations including ``good" diquarks are lighter than those including ``bad" diquarks, 
so the chiral invariant mass for each baryon field should follow $m_0^{\eta} \gtrsim m_0^{\psi} \sim m_0^{\chi}$. 
For simplicity, in this paper,  we set $m_0^{\psi} = m_0^{\chi}$.


\section{Quark diagram for Yukawa interaction}
\label{Yukawa}

In this section we study the mass spectra 
of octet-baryons in a model with first-order Yukawa interactions. The Yukawa interaction contains the coupling to the $\sigma$ fields whose 
condensation breaks the chiral symmetry.
In the standard linear $\sigma$ model nucleon masses arise from the $\sigma $ condensate.

For writing the Yukawa interactions,
we introduce a $3 \times 3$ matrix field $M$ expressing a nonet of scalar and pseudoscalar mesons made of a quark and an anti-quark. 
The representation under SU(3)$_{L} \times $ SU(3)$_{R}$ is 
\begin{align}
M \sim (3, \bar{3}) \,.
\end{align}
We can then construct the chiral invariant Yukawa interaction terms at the first order of $M$ for several combinations of $\psi, \psi^{{\rm mir}},  \chi, \chi^{{\rm mir}}, \eta, \eta^{{\rm mir}} $ fields. 
 In the following, we list the possible Yukawa interactions at first order using a way based on the quark diagrams developed in Ref.~\cite{Minamikawa:2023ypn}.

\begin{figure}
  \centering
  \includegraphics[scale=0.35]{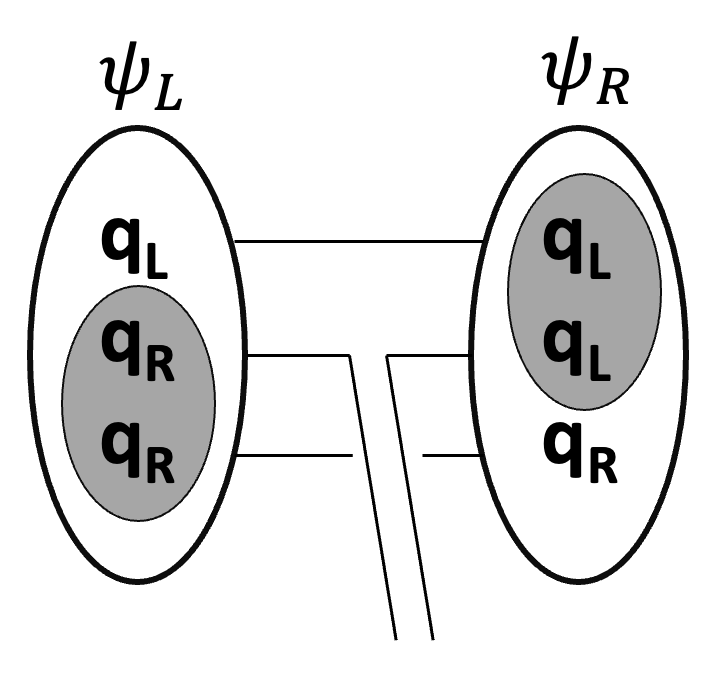}
  \caption{Yukawa  interaction   between $(3, \bar{3})$ and $(\bar{3}, 3)$ baryon fields.
  }
  \label{Yukawa33}
\end{figure}

We first consider the Yukawa interaction between $(3, \bar{3}) + (\bar{3}, 3)$ representations.
We show the corresponding quark diagram in Fig.~\ref{Yukawa33}.
In this figure, the scalar field $M$ couples to a quark $q_R$ 
in the ``good'' 
diquark included in the $(3_L, \bar{3}_R)$ representation. After the chiral flipping, the $(\bar{3}_L, 3_R)$ representation is formed. 
The same is true after exchanges of $L$ and $R$. 
We then construct the chiral invariant term at the first order in $M$ as
\begin{align}
\mathcal{L}^{\psi} = &g_{1} \left(\epsilon_{abc}\epsilon^{\alpha\beta\sigma}(\bar{\psi}_{R})^{a}_{\alpha}(M)^{b}_{\beta}(\psi_{L})^{c}_{\sigma} + h.c \right) \nonumber \\
&+g_2\left(\epsilon^{abc}\epsilon_{\alpha\beta\sigma}(\bar{\psi}^{{\rm mir}}_{R})_{a}^{\alpha}(M)_{b}^{\beta}(\psi^{{\rm mir}}_{L})_{c}^{\sigma} +  h.c \right) \ ,
\end{align}
where the $\epsilon_{ijk}$ is the totally asymmetric tensor.

\begin{figure}
  \centering
  \includegraphics[scale=0.35]{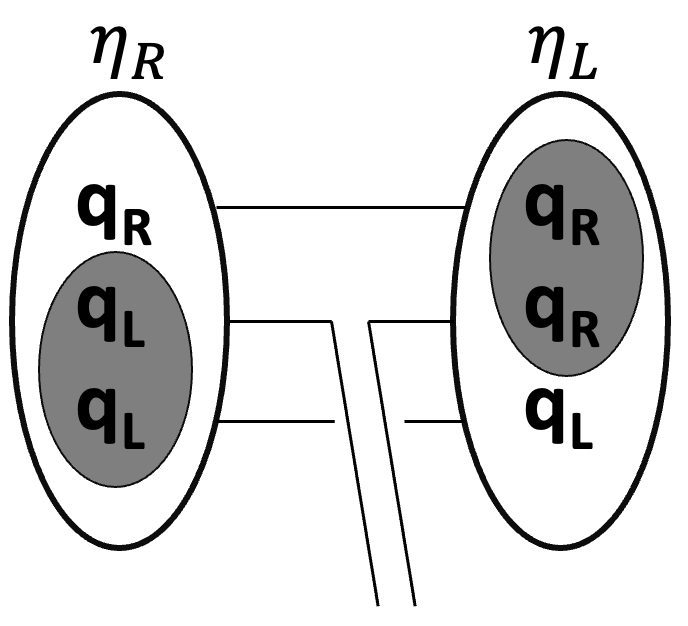}
  \caption{Yukawa  interaction couplings   between $(3, 6)$ and $(6, 3)$ baryon fields.
  }
  \label{Yukawa36_63}
\end{figure}

We next consider the case only with $(3, 6) + (6, 3)$ representation. 
The corresponding quark diagram is shown in Fig.~\ref{Yukawa36_63}.
As shown in this figure, the $M$ field  
couples to the a quark in the ``bad'' diquark included in the 
$(6_{L}, 3_{R})$ representation, forming another ``bad'' diquark in the $(3_{L}, 6_{R})$ representation.
The resultant 
Lagrangian is  written as
\begin{align}
\mathcal{L}^{\eta} = &g_3 \left( (\bar{\eta}_{1r})_{(ab, \alpha)} (M)^{a}_{\beta} (\eta_{1l})^{(b, \alpha\beta)}  + h.c\right) \nonumber \\
&g_4 \left( (\bar{\eta}^{\mir}_{r})_{(a, \alpha\beta)} (M^{\dagger})^{\alpha}_{b} (\eta^{\mir}_{l})^{(ab, \beta)}  + h.c\right).
\end{align}

\begin{figure}
  \centering
  \includegraphics[scale=0.35]{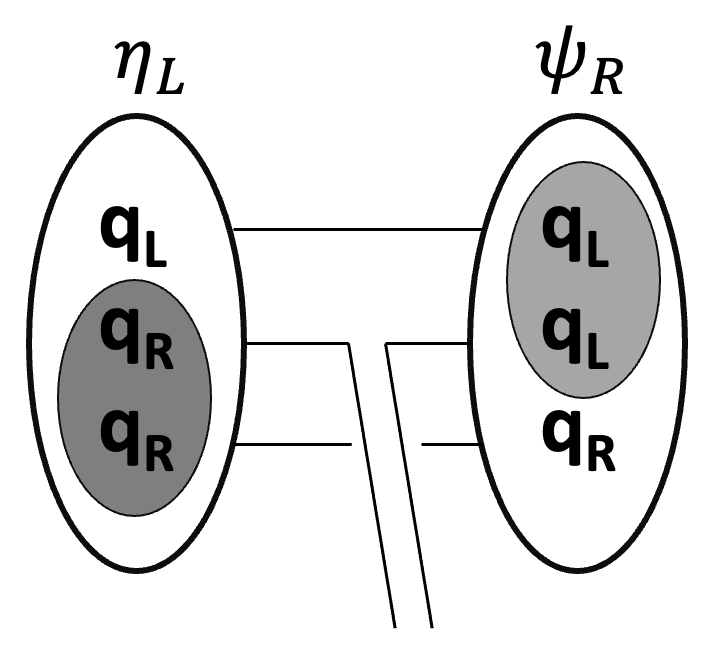}
  \caption{Yukawa interaction  
  between  $(3,6)$  
  and $(\bar{3}, 3)$ baryon fields. 
   }
  \label{Yukawa33_36}
\end{figure}

\begin{figure}
  \centering
  \includegraphics[scale=0.35]{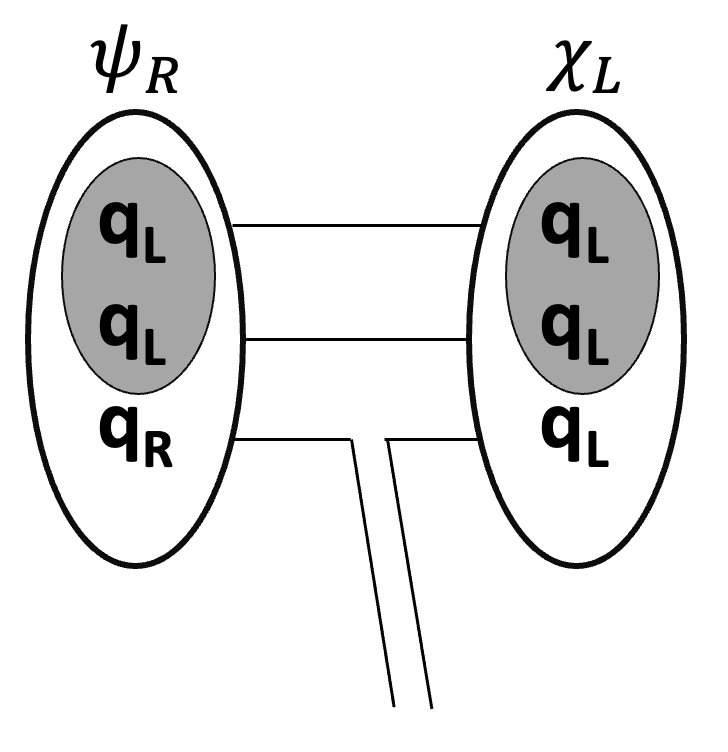}
  \caption{Yukawa  interaction  
  between $(\bar{3}, 3)$ and $(8,1)$  
  baryon fields.
  }
  \label{Yukawa33_18}
\end{figure}

For the Yukawa interactions between $\psi$ and $\eta$, the matrix $M$ couples to a quark $q_L$ in the ``good'' diquark included in the 
$(\bar{3}_L, 3_R)$ representation.
After the chiral flipping the $(3, 6)$ representation is formed as shown in Fig.~\ref{Yukawa33_36}.
The  
chiral invariant Lagrangian at the leading order in $M$ is constructed as 
\begin{align}
\mathcal{L}^{\psi\eta} = &y_1\left[\epsilon_{abc}(\bar{\psi}_{R})^{a}_{\alpha}(M)^{b}_{\beta}(\eta_L)^{(c, \alpha\beta)} + h.c \right] \nonumber\\
& +  y_3\left[\epsilon_{\alpha\beta\sigma}(\bar{\psi}^{{\rm mir}}_{R})^{a}_{\alpha}(M)^{b}_{\beta}(\eta_L)^{(c, \alpha\beta)} + h.c \right] \ .
\end{align}
For the Yukawa interactions between $\psi$ and $\chi$, a spectator quark $q_L$ in $(3_L, \bar{3}_R)$ representation couples to $M$ and flips the chirality  to form the representation $(1_L, 8_R)$ as shown in  Fig.~\ref{Yukawa33_18}. 
The relevant Lagrangian is written as
\begin{align}
\mathcal{L}^{\psi\chi} = &y_{2} {\rm tr} \left(  \bar{\psi}_{R}M^{\dagger}\chi_{L} + h.c \right) \nonumber \\
&+y_{4} {\rm tr} \left( \bar{\psi}_{R}^{{\rm mir}} M \chi_{L}^{{\rm mir}} + h.c \right) \ .
\end{align}
Here we have to emphasize that, at the first orders in $M$, 
there are no Yukawa interactions that couple $\chi_{L}$ and $\chi_{R}$ fields. 
This is because the $\chi$ field contains three valence quarks with all left-handed or right-handed 
so that Yukawa interactions with $\chi$ should include three quark exchanges that flip the chirality of three quarks.

Finally, we consider the Yukawa interaction between $(3_{L}, 6_{R})$ and $(1_{L}, 8_{R})$ representations. 
\begin{figure}
  \centering
  \includegraphics[scale=0.38]{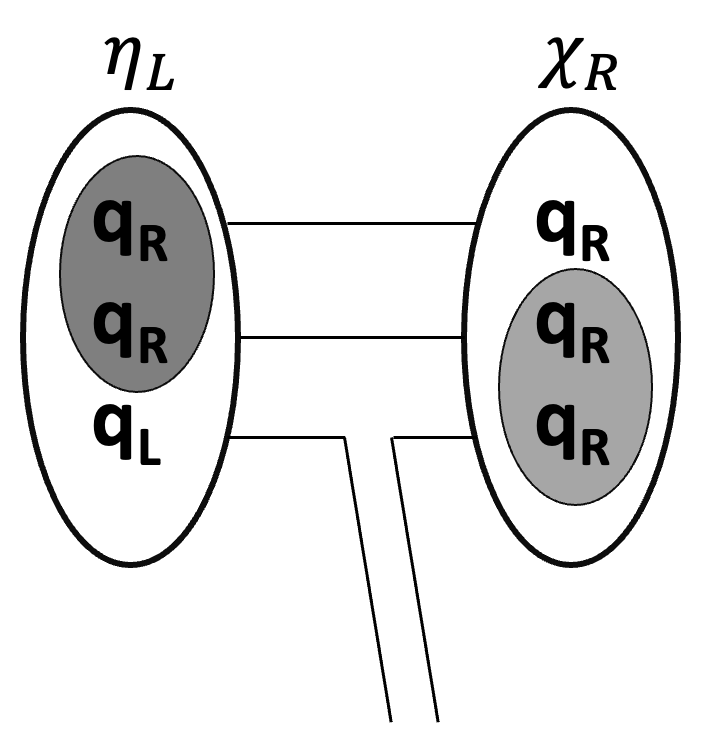}
  \caption{Yukawa couplings between $(3, 6)$ and $(1,8)$  
  baryon fields.
 }
  \label{Yukawa36_18}
\end{figure}
The relavant quark diagram is shown in Fig.~\ref{Yukawa36_18} and the Lagrangian is written as  
\begin{align}
\mathcal{L}^{\eta\chi} = &y_{5} \left(\epsilon^{\alpha\sigma\rho}(\bar{\eta}_{l})_{(a, \alpha\beta)} (M)^{a}_{\sigma} (\chi_{r})^{\beta}_{\rho} + h.c\right) \nonumber \\
& y_6 \left(\epsilon^{acd}(\bar{\eta}^{\mir}_{l})_{(ab, \alpha)} (M^{\dagger})^{\alpha}_{c} (\chi^{\mir}_{r})^{b}_{d} + h.c\right).
\end{align}

Now, the total Lagrangian for the Yukawa interactions at first order in $M$ is expressed as  
\begin{align}
\mathcal{L} = \mathcal{L}^{\psi} + \mathcal{L}^{\eta} + \mathcal{L}^{\psi\eta} + \mathcal{L}^{\psi\chi} + \mathcal{L}^{\eta\chi} + \mathcal{L}_{m_0} \,.
\end{align}
 We should note that the Lagrangian for the Yukawa interactions completely agree with the one provided in Ref.~\cite{Nishihara:2015fka}. 

\begin{table}[ht]
\caption{
Physical inputs of the decay constants for pion and kaon
\cite{Workman:2022ynf},
and the VEV of the meson field  $\langle M \rangle={\rm diag}\{{\alpha,\alpha,\gamma}\}$. 
}
\label{tab-condensate-input}
\centering
\begin{tabular}{c|c}
\hline\hline
$f_\pi$ & 93 MeV \\
$f_K$ & 110 MeV \\
$2\alpha$ & $f_\pi(=93\,\mathrm{MeV})$ \\
$2\gamma$ & $2f_K-f_\pi(=127\,\mathrm{MeV})$ \\
\hline\hline
\end{tabular}
\end{table}

\section{Numerical fitting}
\label{Numerical}

\begin{table*}[htb]
\caption{
Physical inputs for the baryon masses belonging to four SU(3)-flavor octets.  
}
\label{tab-mass-inputs}
\centering
\begin{tabular}{c||c|c|c|c}
 & \multicolumn{4}{c||}{Mass inputs for octet members [MeV]} \\
\hline\hline
$J^P$ & $N$ & $\Lambda$ & $\Sigma$ & $\Xi$  \\
\hline
$m_1: 1/2^+$(G.S.) & 
$N(939)$: $939$ & 
$\Lambda(1116)$: $1116$ & 
$\Sigma(1193)$: $1193$ & 
$\Xi(1318)$: $1318$  \\
$m_2: 1/2^+$ & 
$N(1440)$: $1440$ & 
$\Lambda(1600)$: $1600$ & 
$\Sigma(1660)$: $1660$ & 
$\Xi(?)$:   \\
$m_3: 1/2^-$ & 
$N(1535)$: $1530$ & 
$\Lambda(1670)$: $1674$ & 
$\Sigma(1750)$: $1750$& 
$\Xi(?)$:   \\
$m_4: 1/2^-$ & 
$N(1650)$: $1650$ & 
$\Lambda(1800)$: $1800$ & 
$\Sigma(?)$: & 
$\Xi(?)$:  \\
\hline
\end{tabular}
\end{table*}

In the previous section, we have constructed the Yukawa interactions at the first order in the scalar $M$ field based on the quark-line diagrams. 
In this section, we 
fit the Yukawa couplings to the existing mass spectra of baryons.

We take the mean field approximation  in which the meson field $M$ is replaced with its mean field as  
$\langle M \rangle = {\rm diag}(\alpha, \beta, \gamma)$.
 In the following anaysis we assume the isospin symmetry by taking 
$\alpha = \beta$. It is convenient to introduce a unified notation for the chiral representations of baryons as $\Psi_{i} = (\psi_i, \eta_i, \chi_i, \gamma_5\psi^{{\rm mir}}_i, \gamma_5\eta_i^{{\rm mir}}, \gamma_5\chi_i^{{\rm mir}})^{T}$ with $i = N, \Lambda, \Sigma, \Xi$.
We then calculate the mass terms of baryons in the form of 
\begin{align}
\mathcal{L} =  \sum_{i=N, \Lambda, \Sigma, \Xi} \bar{\Psi}_{i} \hat{M}_{i}\Psi_{i} \ , 
\end{align}
where the  $\hat{M}_{i} (i = N, \Lambda, \Sigma, \Xi )$ are the mass matrices for baryons. As an example, the mass matrix for nucleon is 
\begin{equation}
M_N=\left(\begin{array}{cccccc}
g_1 \alpha & -\frac{3 y_1}{\sqrt{6}} \alpha & -y_2 \alpha & m_0^{\psi} & 0 & 0 \\
& \frac{g_3}{2} \alpha & -\frac{3 y_5}{\sqrt{6}} \alpha & 0 & m_0^{\eta} & 0 \\
& & 0 & 0 & 0 & m_0^{\chi} \\
& & & -g_2 \alpha & \frac{3 y_3}{\sqrt{6}} \alpha & y_4 \alpha \\
& & & & -\frac{g_4}{2} \alpha & \frac{3 y_6}{\sqrt{6}} \alpha \\
& & & & & 0
\end{array}\right) \ ,
\end{equation}
where we omit the lower triangular part of the matrix $M_N$, which is understood from the fact that $M_N$ is the symmetric matrix $\left(M_N\right)^T= M_N$. 

Diagonalizing this $6 \times 6$ matrix $\hat{M}_{i}$, we obtain six mass eigenvalues. 
We focus on the first four states of baryons in this research as the remaining two states completely come from predictions with large ambiguity. 
We determine the vacuum expectation values (VEVs) of the meson field $M$ from the decay constants of pion and kaon as
\begin{align}
2\alpha = f_{\pi}, \quad 2\gamma = 2f_K - f_{\pi} \,.
\end{align}
In Table~\ref{tab-condensate-input}, the input values of $f_{\pi}$ and $f_K$ are shown together with the determined values of $\alpha$ and $\gamma$.

%

%
\begin{figure*}\centering
\begin{subfigure}{0.4\hsize}\centering
	\includegraphics[width= 0.95\hsize]{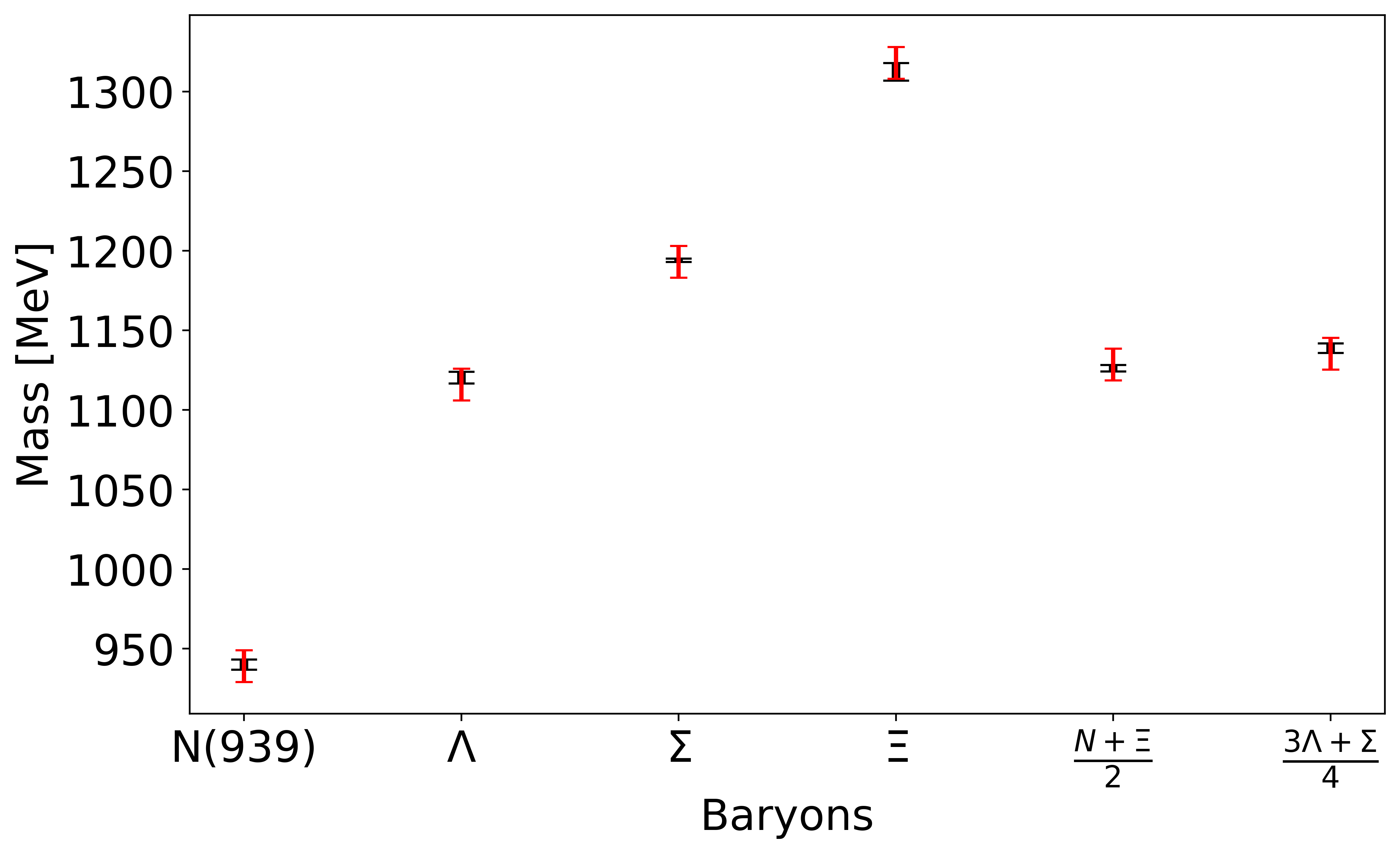}
	\caption{}
\end{subfigure}
\begin{subfigure}{0.4\hsize}\centering
	\includegraphics[width= 0.95\hsize]{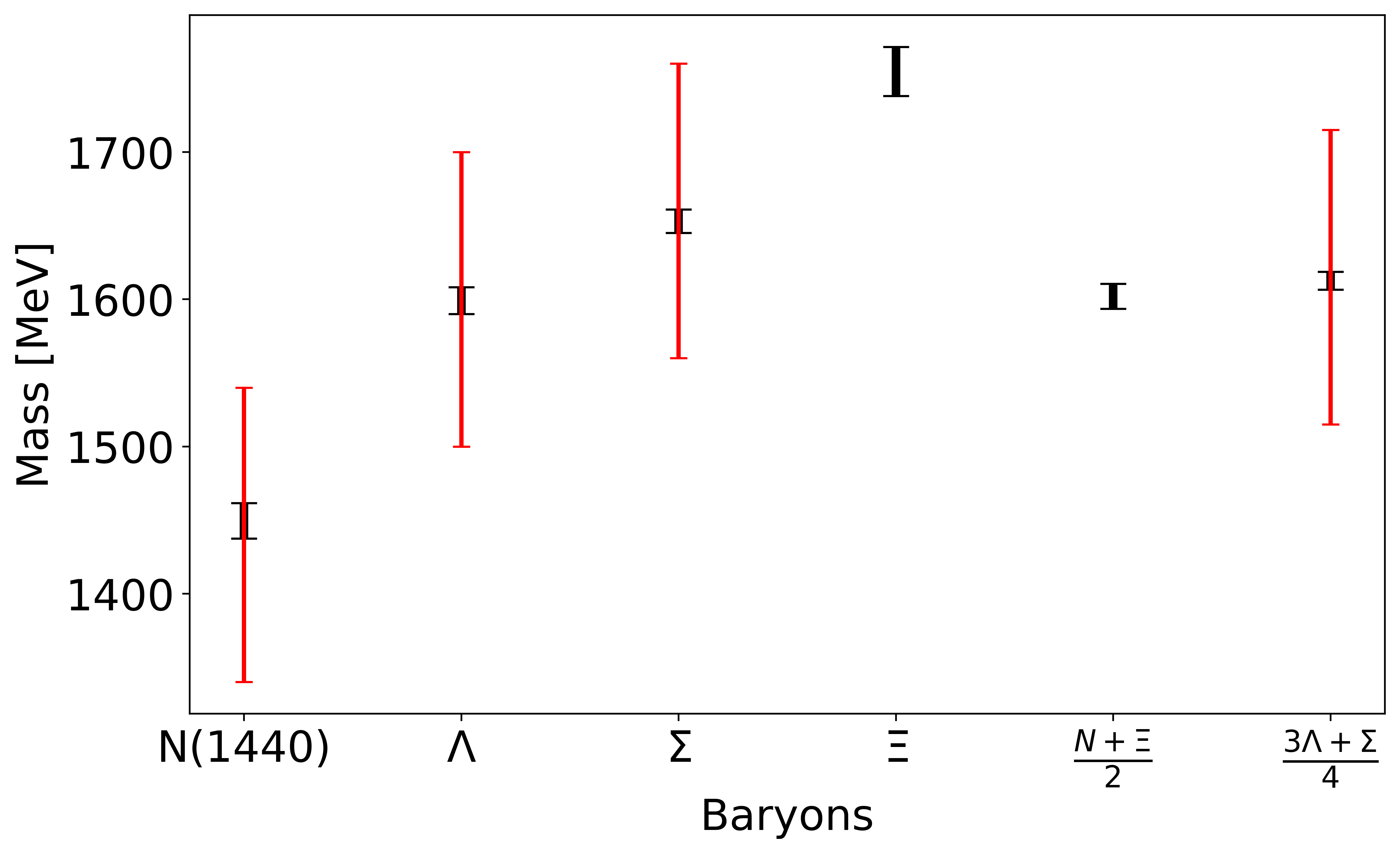}
	\caption{}
\end{subfigure}
\begin{subfigure}{0.4\hsize}\centering
	\includegraphics[width= 0.95\hsize]{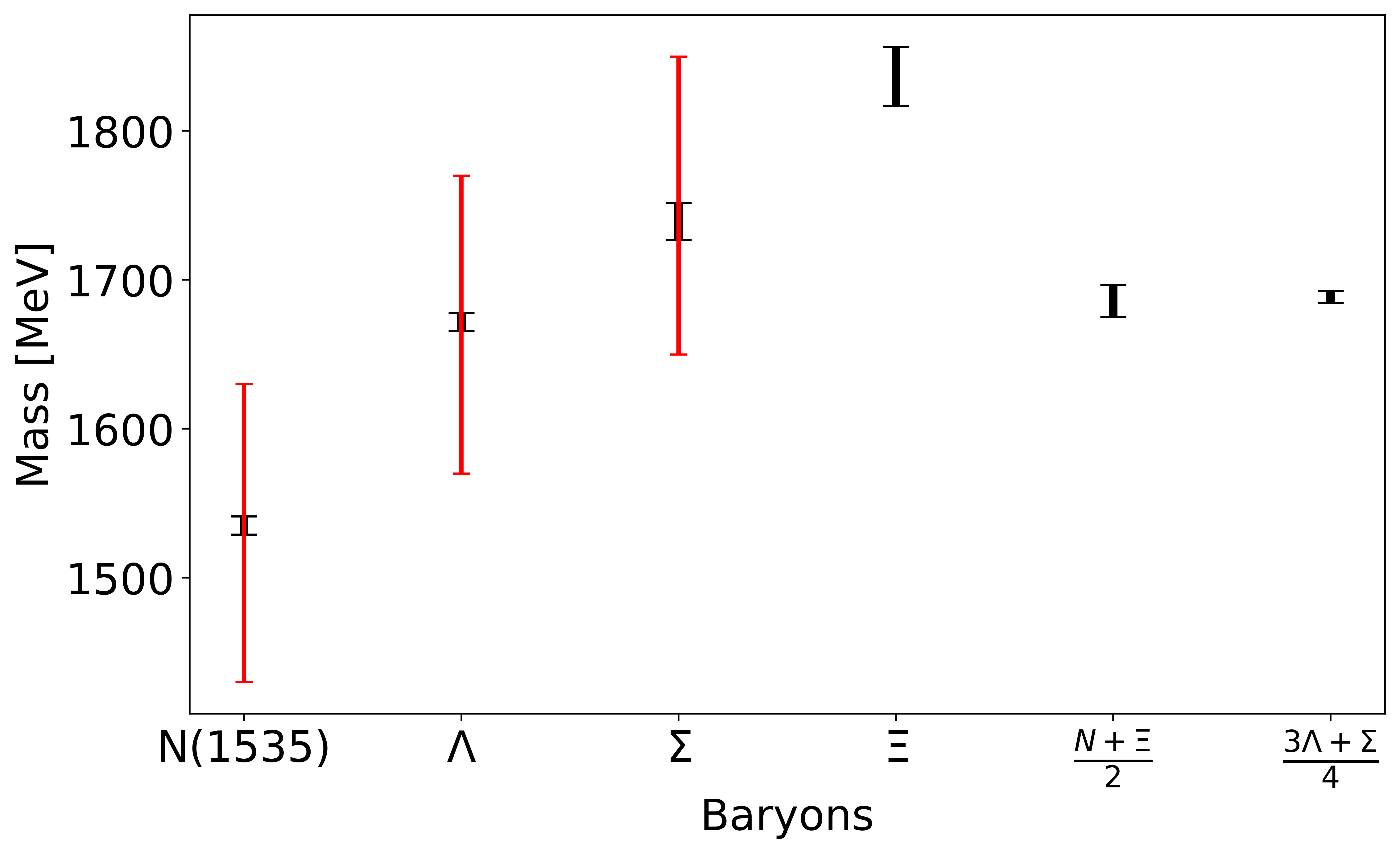}
	\caption{}
\end{subfigure}
\begin{subfigure}{0.4\hsize}\centering
	\includegraphics[width= 0.95\hsize]{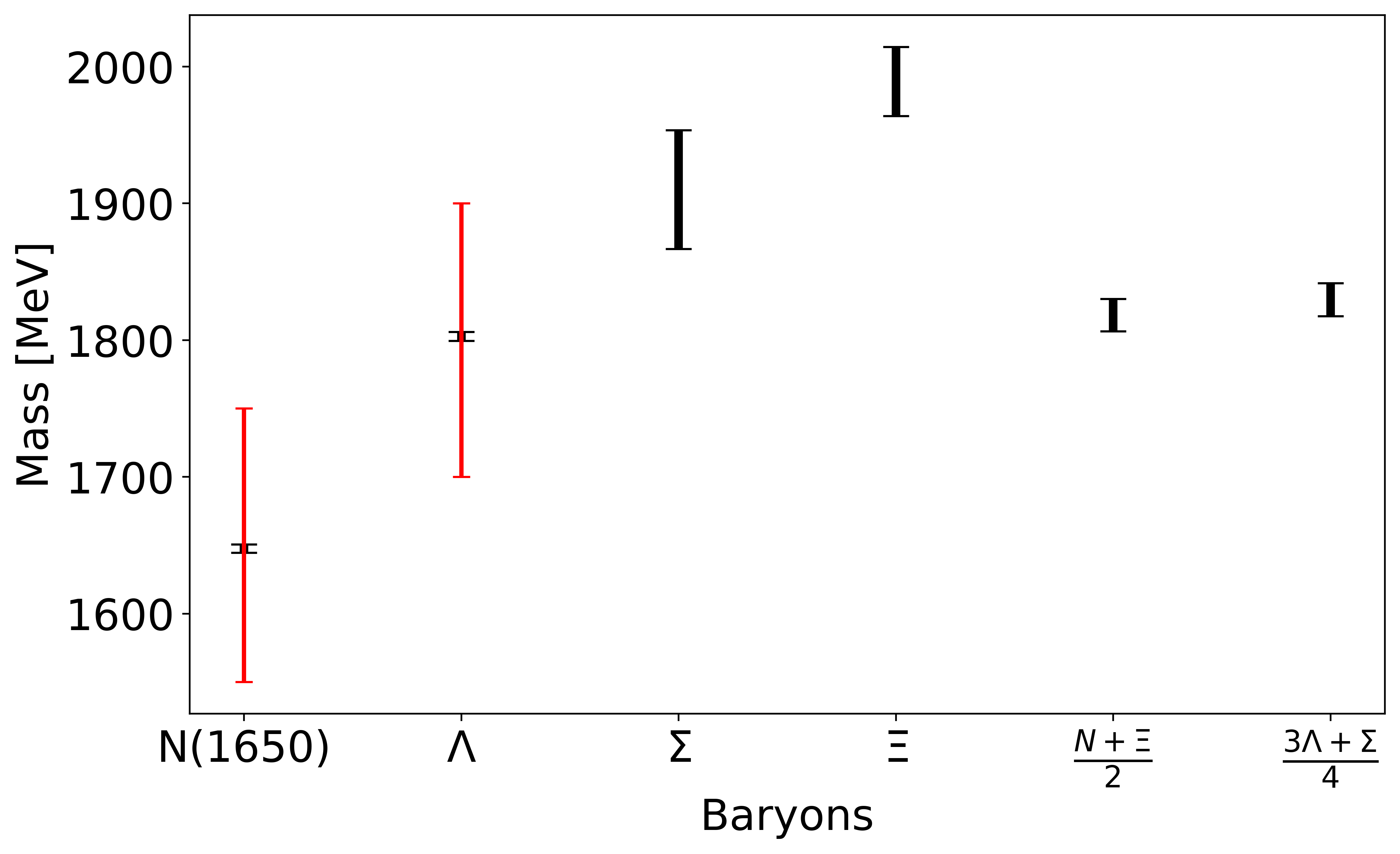}
	\caption{}
\end{subfigure}
\begin{subfigure}{0.4\hsize}\centering
	\includegraphics[width= 0.95\hsize]{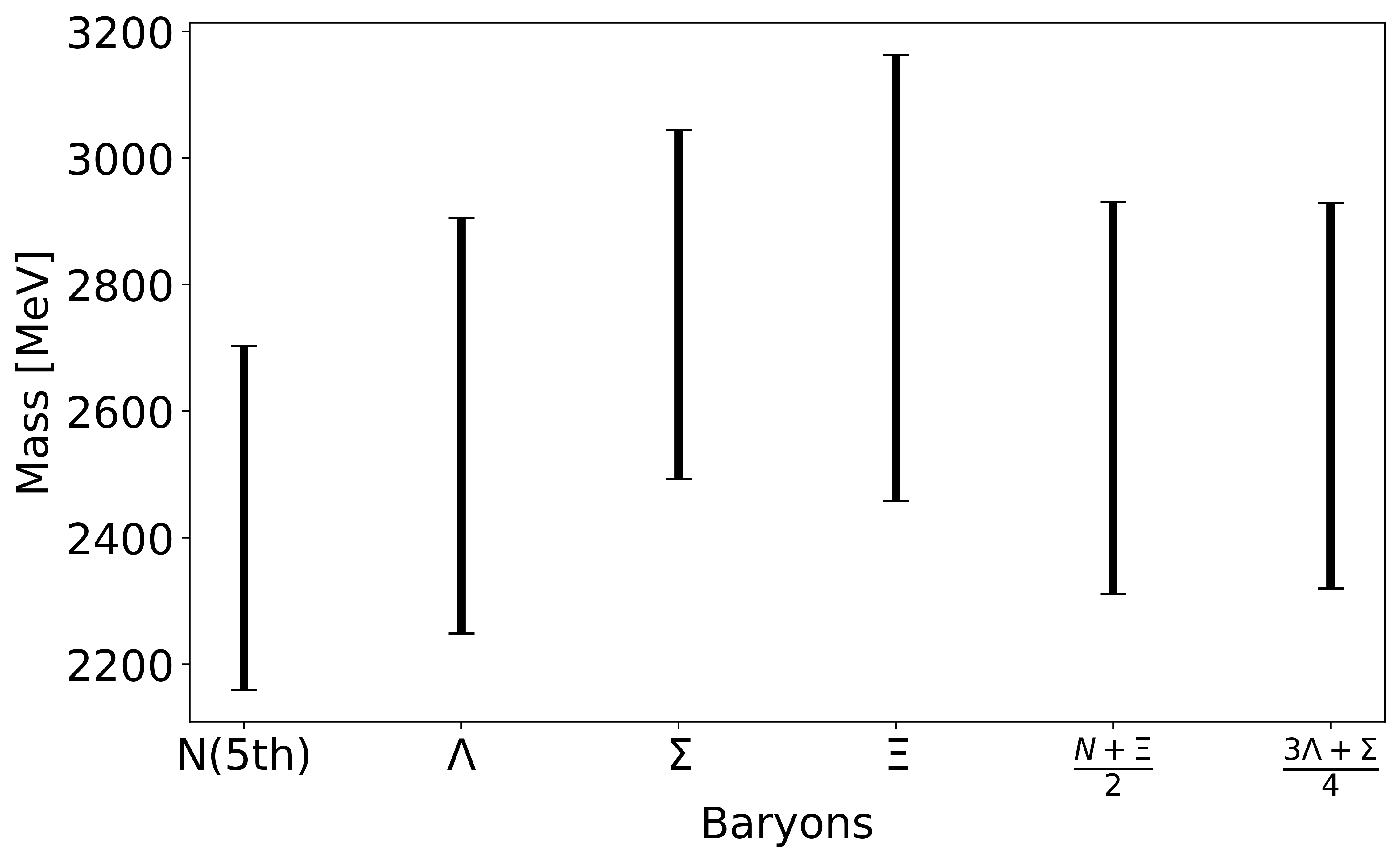}
	\caption{}
\end{subfigure}
\begin{subfigure}{0.4\hsize}\centering
	\includegraphics[width= 0.95\hsize]{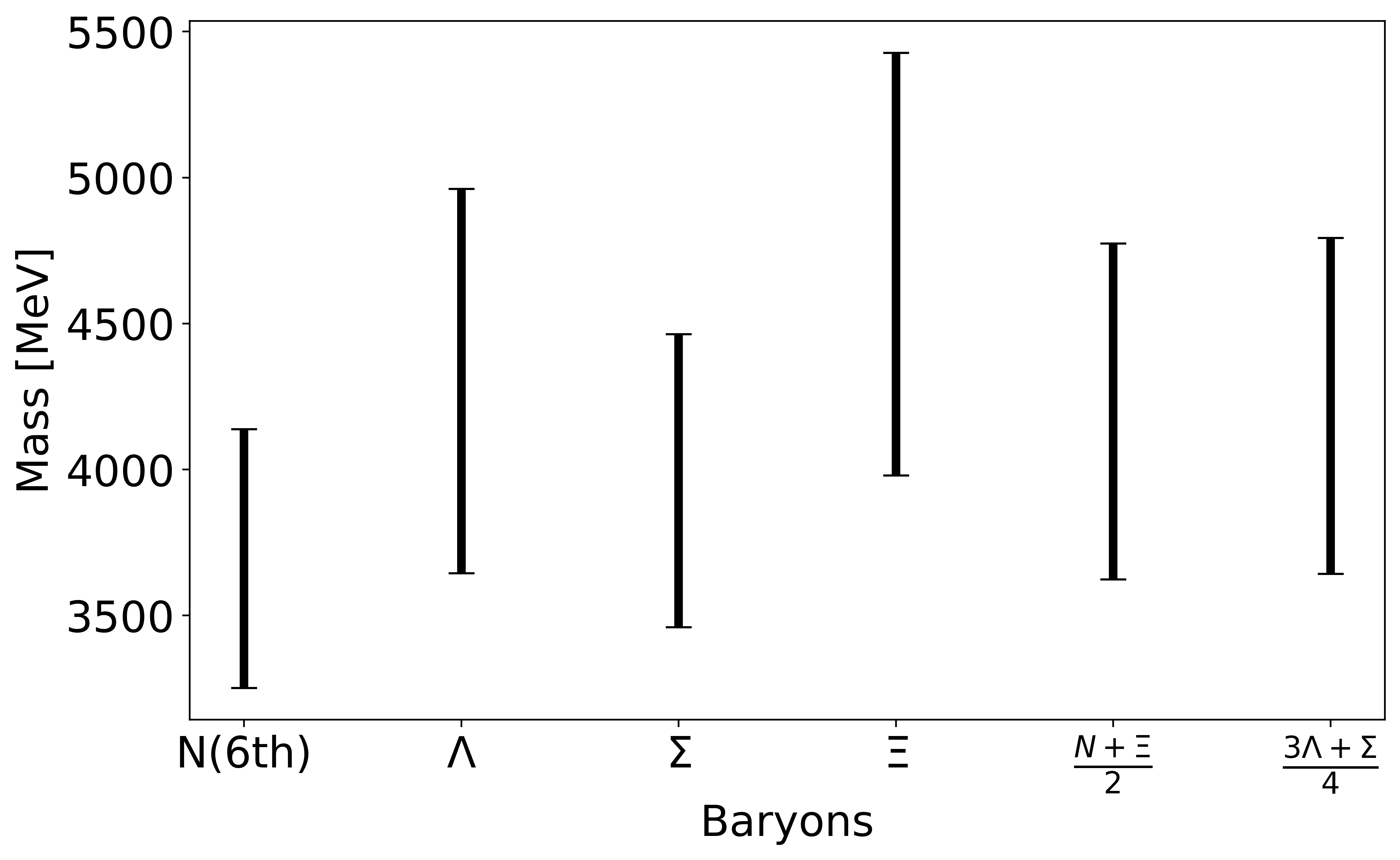}
	\caption{}
\end{subfigure}
\caption{
Masses of octet baryons 
for the chiral invariant mass $m^{\psi}=m^{\chi}=700$ MeV and  $m^{\eta}=1000$ MeV. 
The red 
lines show input values 
the physical inputs 
with errors and  the black 
lines show solutions satisfying 
line shows for solutions satisfy 
$f_{\rm min} < 1$.
}
\label{fig-massspec}
\end{figure*}
\begin{figure*}\centering
\begin{subfigure}{0.4\hsize}\centering
	\includegraphics[width= 0.95\hsize]{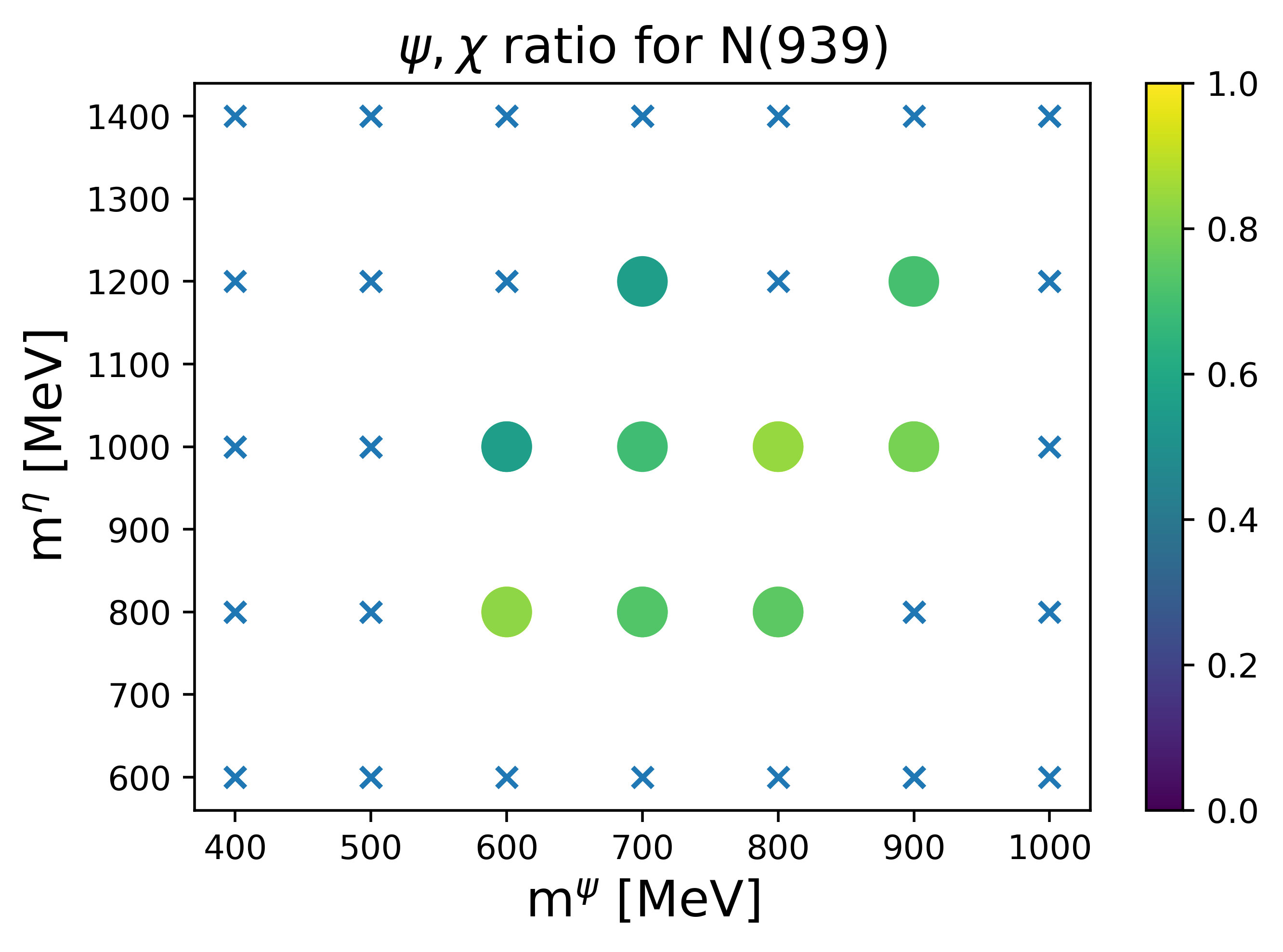}
	\caption{}
	\label{fig-quarklines-spectator-proton}
\end{subfigure}
\begin{subfigure}{0.4\hsize}\centering
	\includegraphics[width= 0.95\hsize]{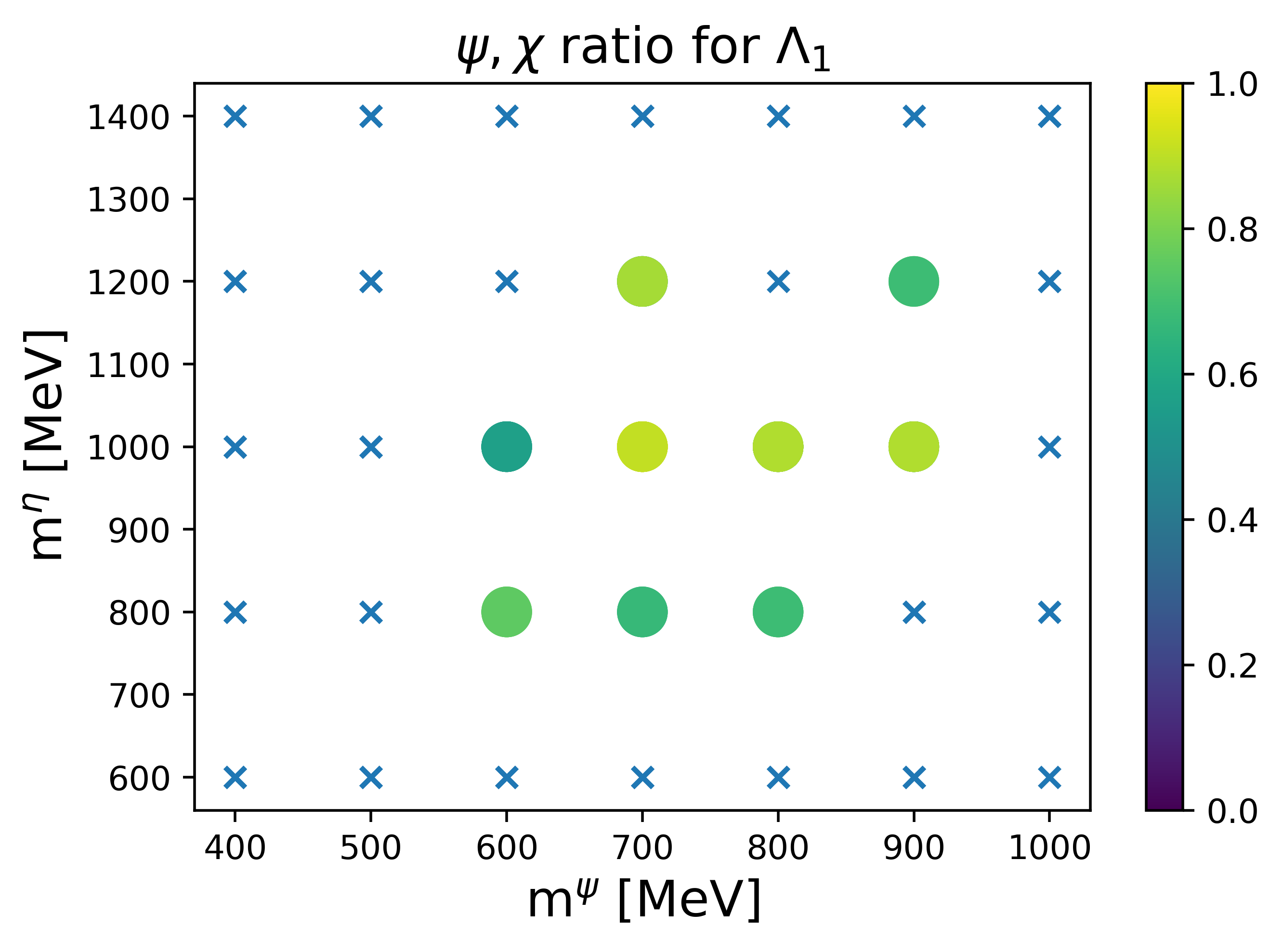}
	\caption{}
	\label{fig-quarklines-spectator-Sigma}
\end{subfigure}
\begin{subfigure}{0.4\hsize}\centering
	\includegraphics[width= 0.95\hsize]{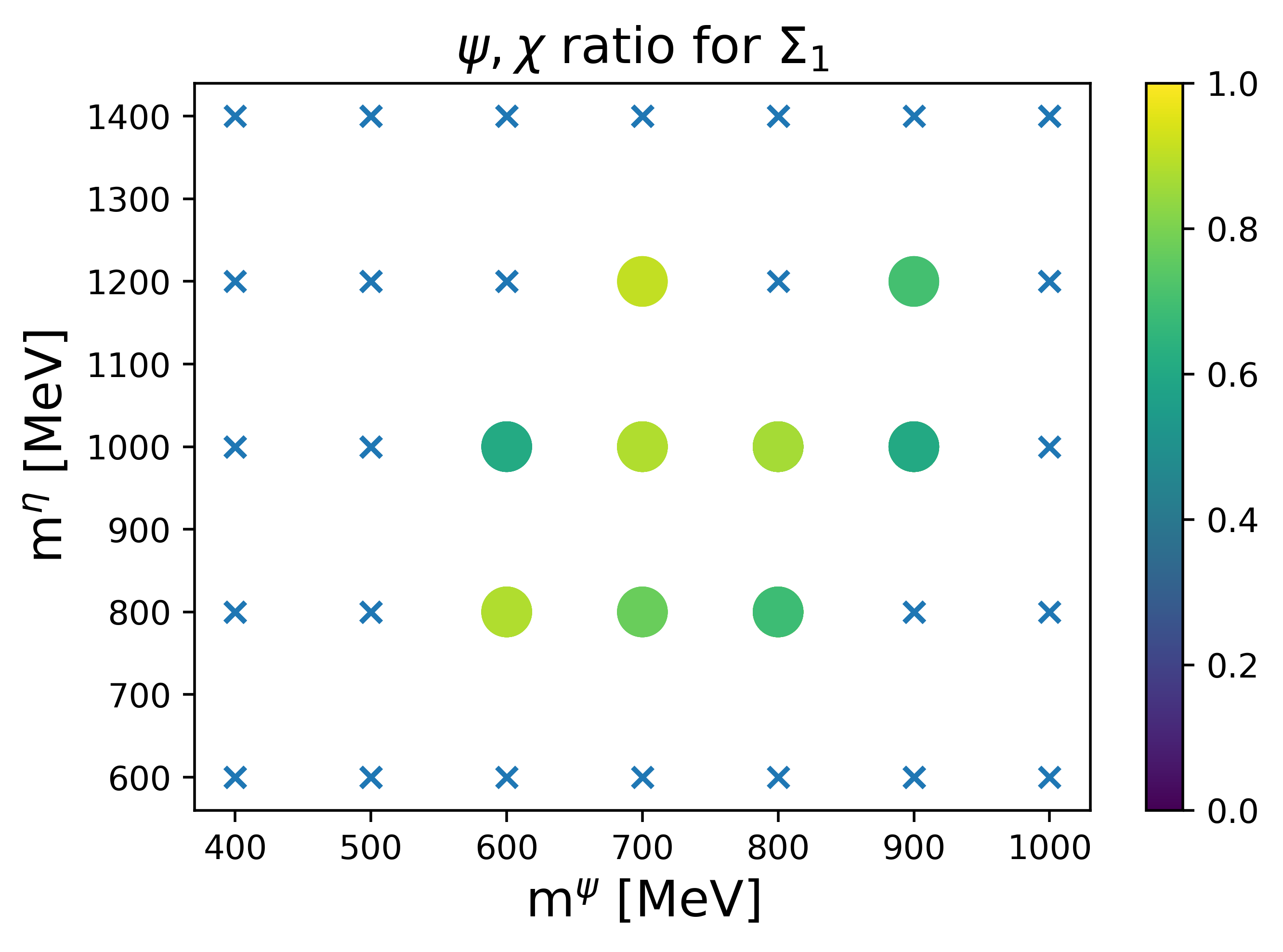}
	\caption{}
	\label{fig-quarklines-spectator-Xi}
\end{subfigure}
\begin{subfigure}{0.4\hsize}\centering
	\includegraphics[width= 0.95\hsize]{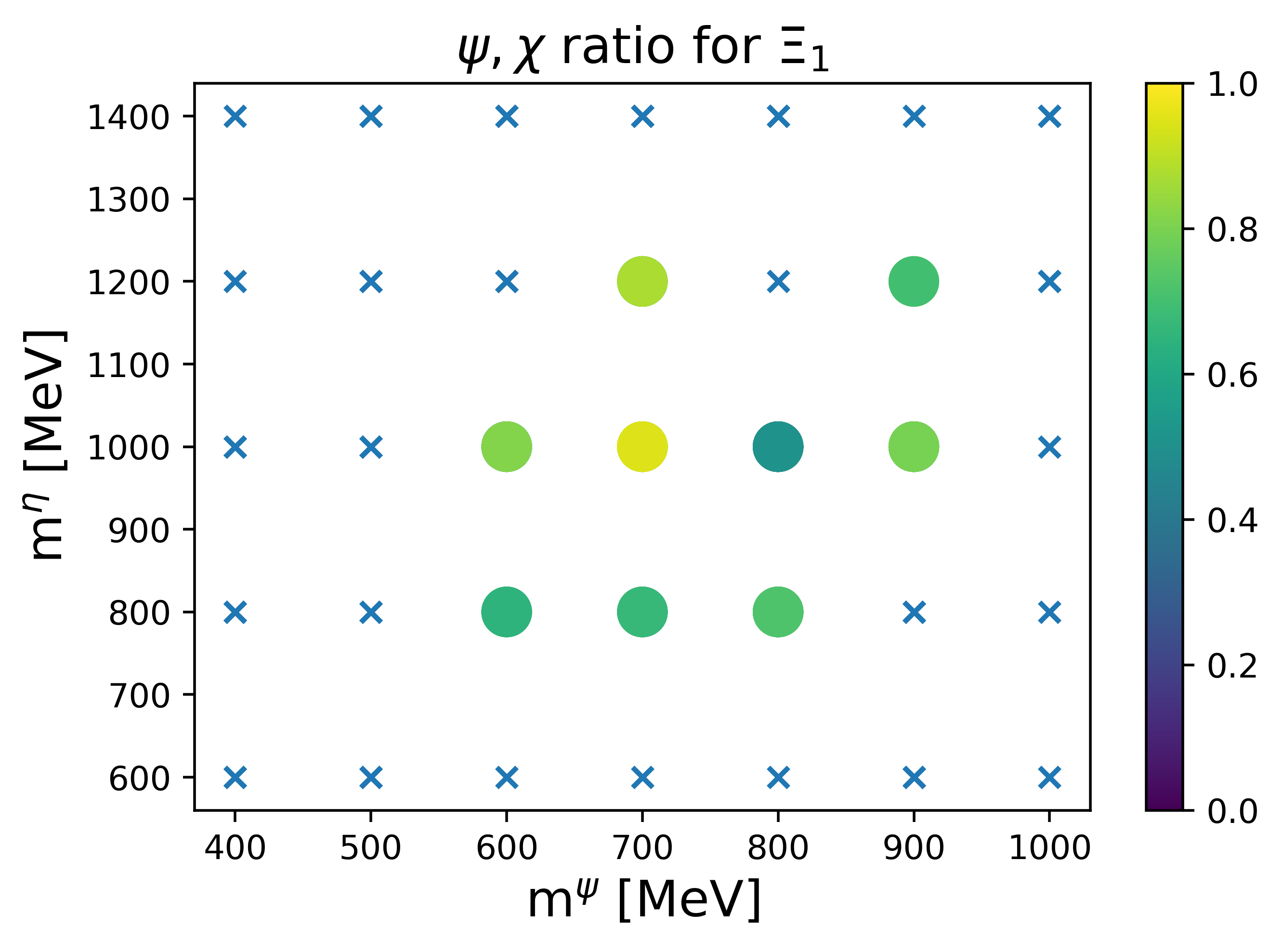}
	\caption{}
	\label{fig-quarklines-spectator-Xi}
\end{subfigure}
\caption{ Numerical results for 
the probability to find representations with ``good'' diquarks,
$ | c_\psi |^2 + | c_{\psi^{\rm mir}} |^2 + | c_\chi |^2 + | c_{\chi^{\rm mir} } |^2 $, 
for each combination of $(m^{\psi(\chi)}, m^{\eta})$.
}
\label{fig-ratio}
\end{figure*}

\begin{figure*}\centering
\begin{subfigure}{0.4\hsize}\centering
	\includegraphics[width= 0.95\hsize]{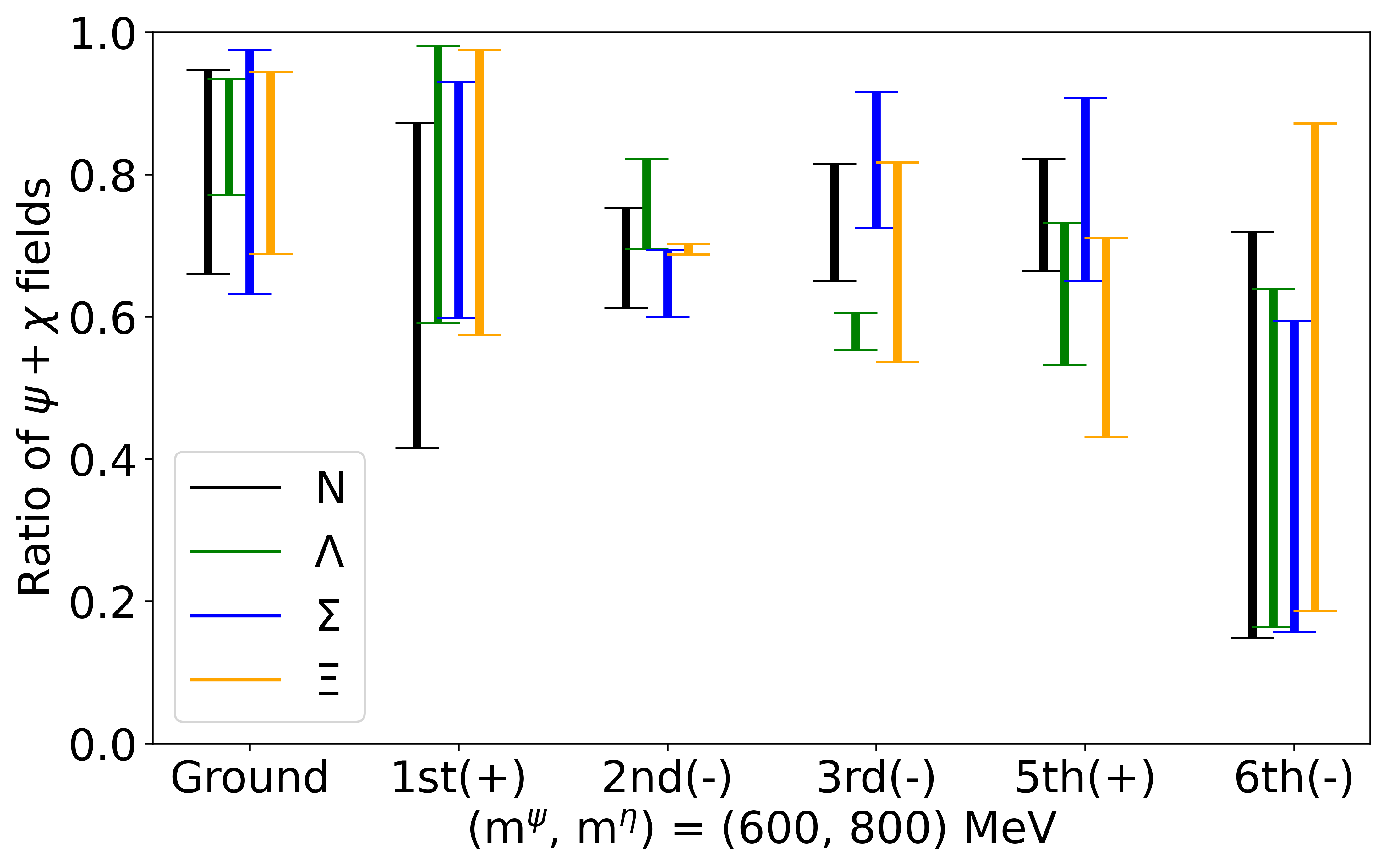}
	\caption{}
\end{subfigure}
\begin{subfigure}{0.4\hsize}\centering
	\includegraphics[width= 0.95\hsize]{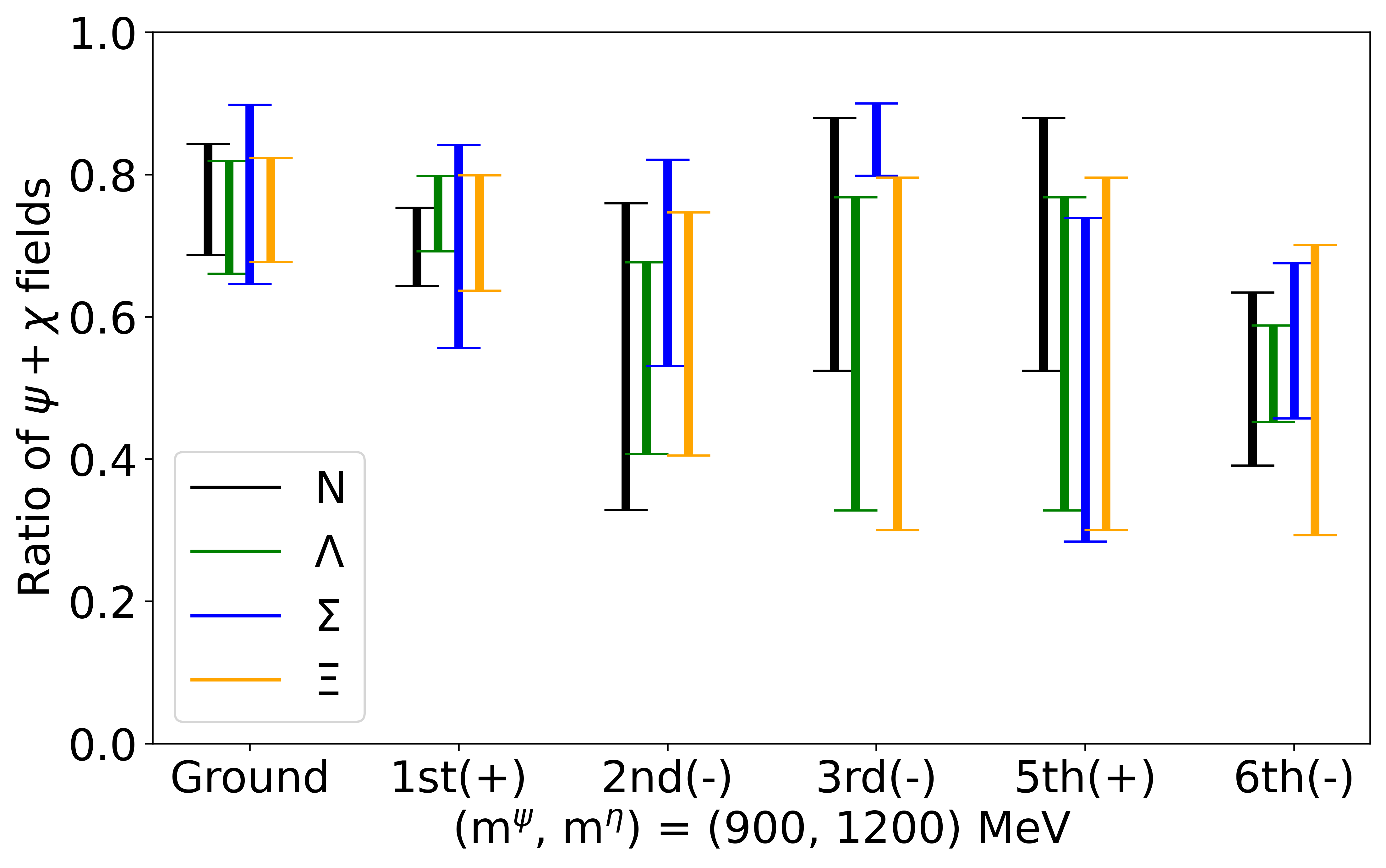}
	\caption{}
\end{subfigure}
\begin{subfigure}{0.4\hsize}\centering
	\includegraphics[width= 0.95\hsize]{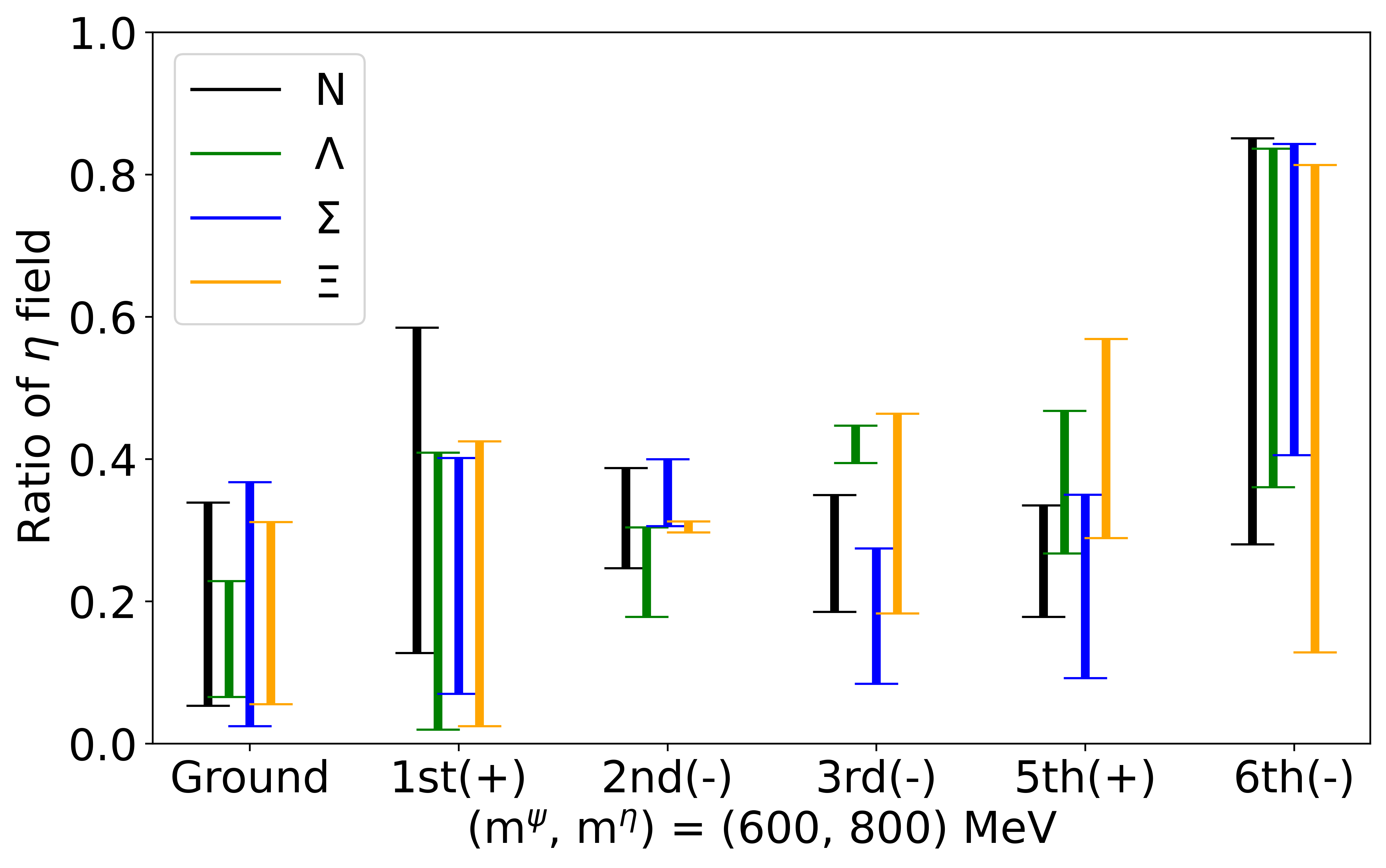}
	\caption{}
\end{subfigure}
\begin{subfigure}{0.4\hsize}\centering
	\includegraphics[width= 0.95\hsize]{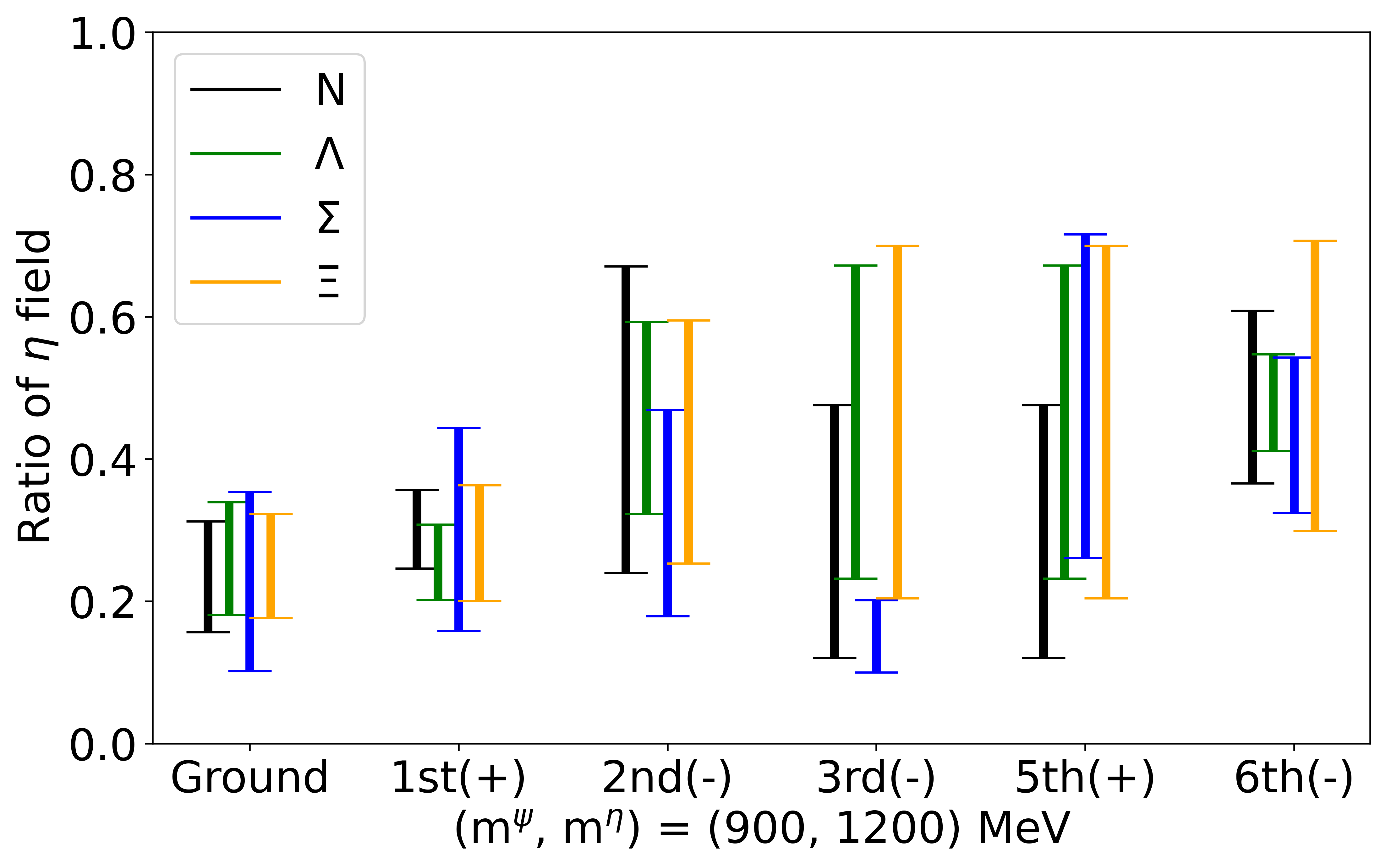}
	\caption{}
\end{subfigure}
\caption{  Numerical results for 
the probability to find representations with ``good'' ((a) and (b)) and ``bad'' ((c) and (d)) diquarks,
$ | c_\psi |^2 + | c_{\psi^{\rm mir}} |^2 + | c_\chi |^2 + | c_{\chi^{\rm mir} } |^2 $ 
and $|c_\eta|^2 + |c_{\eta^{\rm mir} }|^2$ respectively,
for typical choice of $(m^{\psi(\chi)}, m^{\eta})$. 
The sign in the bracket indicates the parity. 
}
%
\label{fig-ratio-excited}
\end{figure*}

\begin{figure*}\centering
\begin{subfigure}{0.4\hsize}\centering
	\includegraphics[width= 0.95\hsize]{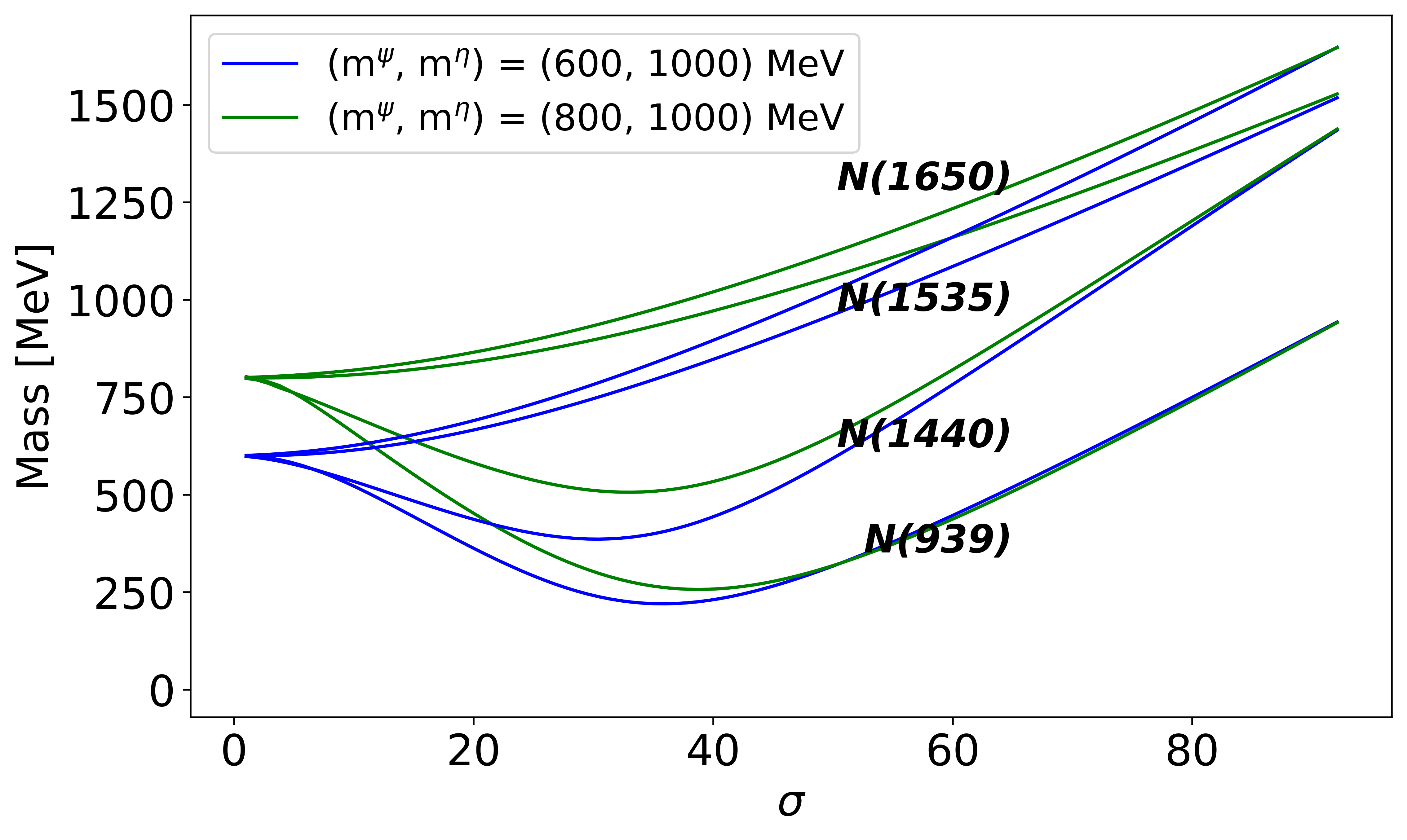}
	\caption{}
\end{subfigure}
\begin{subfigure}{0.4\hsize}\centering
	\includegraphics[width= 0.95\hsize]{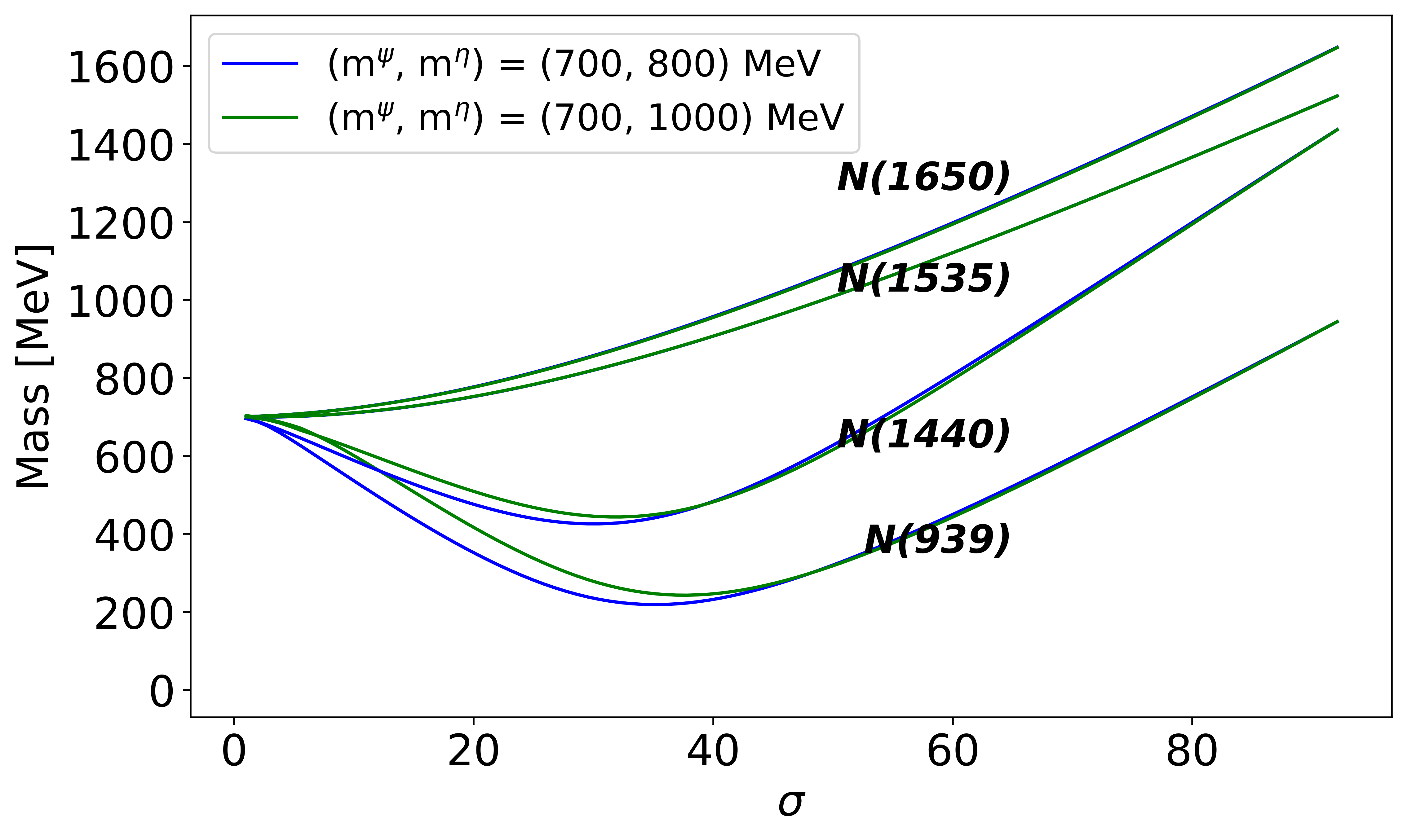}
	\caption{}
\end{subfigure}
\begin{subfigure}{0.4\hsize}\centering
	\includegraphics[width= 0.95\hsize]{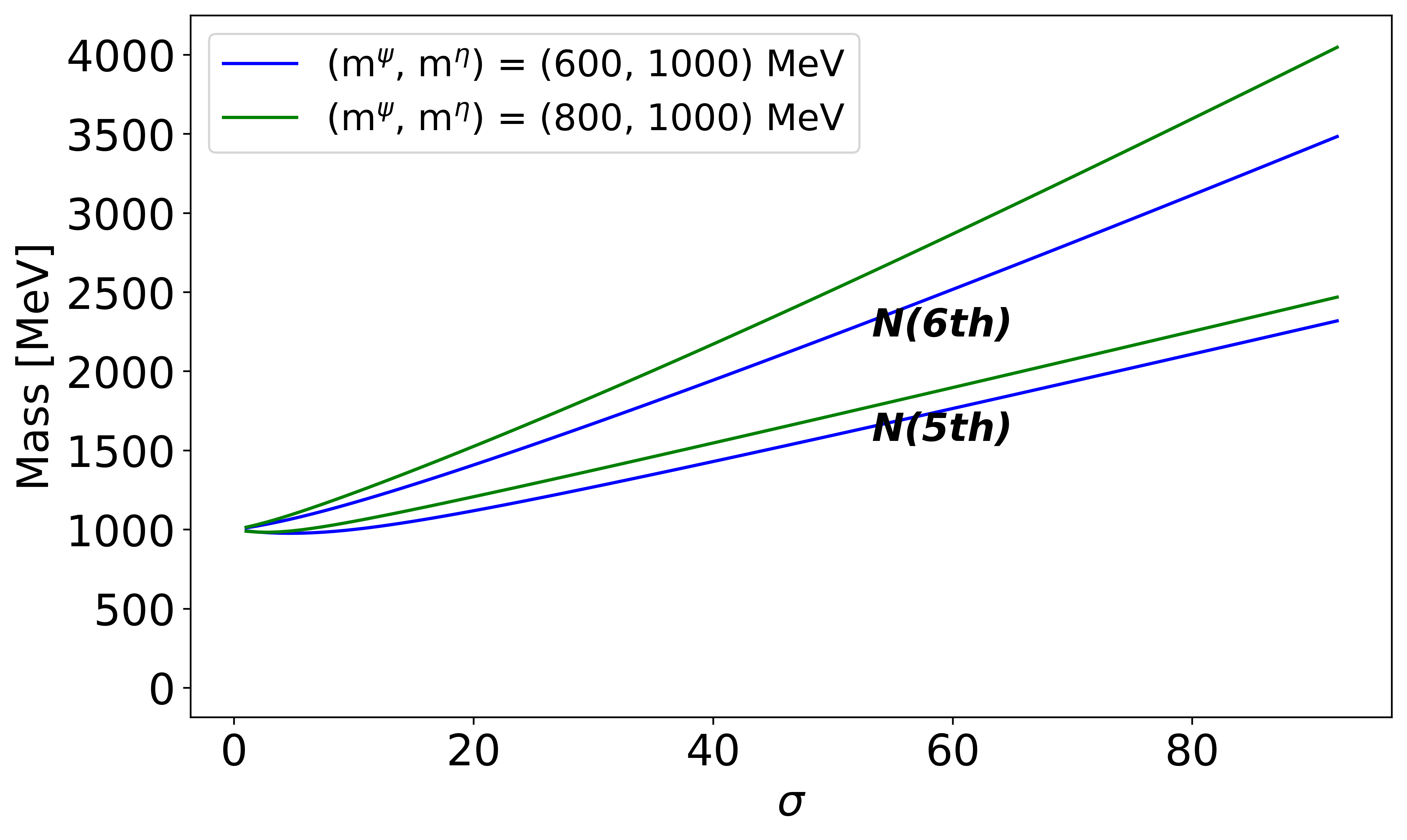}
	\caption{}
\end{subfigure}
\begin{subfigure}{0.4\hsize}\centering
	\includegraphics[width= 0.95\hsize]{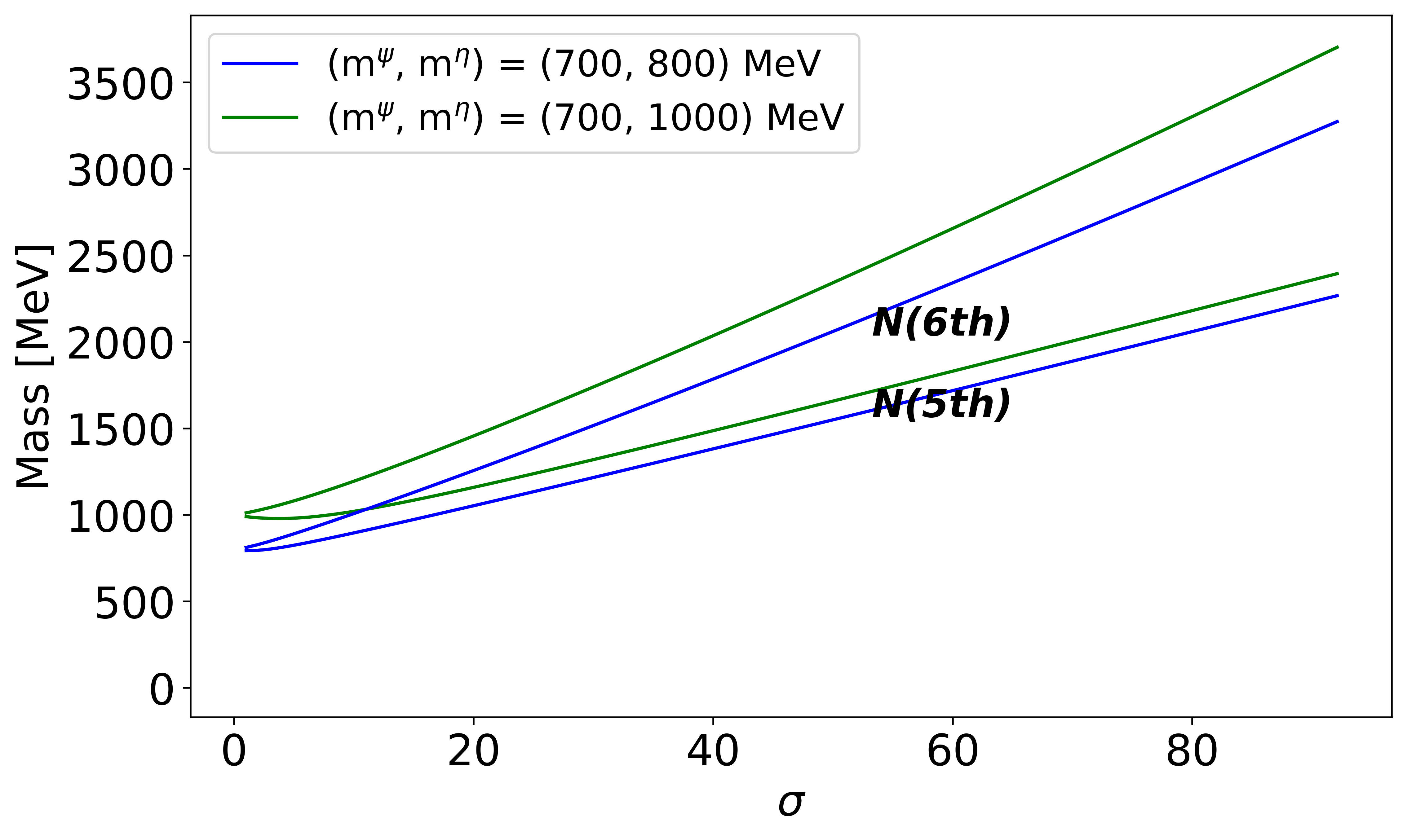}
	\caption{}
\end{subfigure}
\caption{
Numerical results of $\sigma$ dependence of nucleon mass  for each combination of $(m^{\psi(\chi)}, m^{\eta})$. We choose the best fitted value for each combination of ($m^{\psi(\chi)}, m^{\eta}$).
}
\label{N_sigma}
\end{figure*}
%

We determine the model parameters based on the following procedure:
First we fix chiral invariant masses, $m^\psi = m^\chi$ and $m^\eta$ as given and later examine different values of $(m^\psi, m^\eta)$.
Then we have 10 Yukawa couplings $g_{i=1-4}$ and $y_{i=1-6}$ to be used for fitting.
Using twelve mass values in Table~\ref{tab-mass-inputs} as inputs, 
we determine 10 Yukawa coupling constants 
by minimizing the following error function 
\begin{align}
f_{{\rm min}} = \sum_{i=1}^{12} \left(\frac{m_i^{{\rm theory}} - m_i^{{\rm input}}}{\delta m_i} \right)^2 \label{minimize}
\end{align}
where errors  $\delta m_i$ are taken as $\delta m_i = 10$ MeV for the ground-state baryons and $\delta m_i = 100$ MeV for the excited baryons. The 12 baryon states we have chosen are listed in the Particle Data Group (PDG) with reasonable confidence level.
The other 12 states not listed in the PDG are 
the predictions of the present model. 
For these states we demand the Gell-Mann-Okubo mass relation to the following accuracy

%
 %
\begin{align}
\sum_i | \Delta_{{\rm GO}, i} | < 100 \,\, {\rm MeV}
\end{align}
where 
\begin{align}
\Delta_{{\rm GO}, i} = \frac{m_i[N] + m_i[\Xi]}{2} - \frac{3m_i[\Lambda] + m_i[\Sigma]}{4}
\end{align}
with $i$ indicating the octet  generations as in Table~\ref{tab-mass-inputs}. 
This condition reflects the mass difference between up, down and strange quarks
and should remain useful even for the mass difference of excited states.
We simply reject parameter sets which strongly violate this condition.

%

%
In Fig.~\ref{fig-massspec}, we show the resultant  
mass spectrum together with the Gell-Mann-Okubo relation with errors, 
where the error is determined by requiring $ f_{\rm min}^{\rm best} < 1$, for $m^{\psi}=m^{\chi}=700$\,MeV and  $m^{\eta}=1000$\,MeV.  
%
%
The red line shows the experimental values as inputs.
The black line is drawn by accumulating spectra from all parameter sets satisfying $ f_{\rm min}^{\rm best} < 1$
and hence reflects the global aspect of our analyses.  We find that all the baryon masses below 2 GeV are reproduced well. We emphasize that such good fit was not possible in the previous research in Ref.~\cite{Minamikawa:2023ypn}, 
where the $(3, \bar{3}) + (\bar{3}, 3)$ and $(8, 1) + (1, 8)$ representations are used to construct a chiral invariant Lagrangian up to second order Yukawa interactions, while the $(3, 6) + (6, 3)$ representations are assumed to be integrated out. 
In particular the mass hierarchy between $\Sigma$ and $\Xi$ cannot be correctly reproduced, 
indicating the baryon dynamics is not well-saturated by just $(3, \bar{3}) + (\bar{3}, 3 )$ and $(8, 1) + (1, 8 )$ representations. 
In contrast, the current analyses manifestly including the $(6, 3) + (3, 6)$ representation correctly describes the mass hierarchy between excited $\Sigma$ and $\Xi$ states.
This should be natural simply because the $(6, 3) + (3, 6)$ representations with bad diquarks have more correlations with excited states;
attempts to fit excited states without $(6, 3) + (3, 6)$ representations affect fits to the other states,
making 
the 
previous global analyses problematic.
We also make some prediction about the mass of second excited state of $\Sigma$  and excited states of $\Xi$ below 2 GeV. 

%
%

Next, we examine the composition of each representation for a baryon state.
For instance, for a nucleon,
\begin{equation}
|N \rangle = c_\psi |\psi \rangle + c_\chi |\chi \rangle + c_\eta | \eta \rangle + \cdots + c_{\eta^{\rm mir} } | \eta^{\rm mir} \rangle\,, 
\end{equation}
where the flavor wavefunction is normalized to $\langle N | N \rangle = 1$.
We examine the probability to find light representations, 
$ | c_\psi |^2 + | c_{\psi^{\rm mir}} |^2 + | c_\chi |^2 + | c_{\chi^{\rm mir} } |^2 $, 
in a nucleon.
Shown in Fig.~\ref{fig-ratio} by colors are the probability to find $|\psi \rangle$ and $| \chi \rangle$ states  
in the ground states ($N, \Lambda, \Sigma, \Xi$) for given $(m^{\psi}, m^{\eta})$.
The quality of fit (measured by $f_{\rm min}^{\rm best}$) is not always good for some domain of $(m_\psi, m_\chi)$
and such domains with $f^{{\rm best}}_{{\rm min}} \ge 1$ are marked with cross symbols and are omitted from our analyses.
The best fit for given $(m_\psi, m_\chi)$ suggests that
the ground states are well dominated by the $(3, \bar{3}) + (\bar{3}, 3)$ 
and $(8, 1) + (1, 8)$ states including ``good'' diquarks.

We also examine how robust the dominance of 
$(3, \bar{3}) + (\bar{3}, 3)$ 
and $(8, 1) + (1, 8)$ states is by studying the variance around the best fit.
In Fig.~\ref{fig-ratio-excited}, we accumulate the results on composition coming from all the parameter sets satisfying $ f_{\rm min}^{\rm best} < 1$.  
The ground to the fifth excited states are displayed for $(m^\psi, m^\eta) = (600, 800)$ MeV and $(900, 1200)$ MeV.
The results show that the dominance of  $(3, \bar{3}) + (\bar{3}, 3)$ 
and $(8, 1) + (1, 8)$ states is a robust conclusion.
In Fig.~\ref{fig-ratio-excited}, we also examine the fraction of $(6, 3) + (3, 6)$ representation $\eta$.
The overall tendency is that the fraction gently grows with the excitation levels.
These trends are consistent with conventional arguments based on diquark classifications in the hadron spectroscopy.
\begin{table*}
\caption{
$\sigma$ dependence of nucleon masses at vacuum.
}
\label{tab-Yukawa}
\begin{tabular}{c||c|c|c|c}
\hline\hline
	~\quad $(m^{\psi(\chi), m^{\eta}})$ [MeV] ~&~ (600, 1000) ~&~ (800, 1000)~ & ~(700, 800) ~&~ (700, 1000)~  \\
	\hline
	$|\partial m_{N} / \partial \sigma|$ [G.S] & 16.09 & 16.02 & 16.04 & 16.41  \\
	$|\partial m_{N} / \partial \sigma|$ [$N(1440)$] & 20.06 & 19.31 & 19.55 & 19.78  \\
	$|\partial m_{N} / \partial \sigma|$ [$N(1535)$] & 13.99 & 12.23 & 13.16 & 13.16  \\
	$|\partial m_{N} / \partial \sigma|$ [$N(1650)$] & 16.13 & 13.86 & 14.89 & 15.15  \\
	$|\partial m_{N} / \partial \sigma|$ [$N(5{\rm th})$] & 17.43 & 17.90 & 17.21 & 17.70  \\
	$|\partial m_{N} / \partial \sigma|$ [$N(6{\rm th})$] & 30.72 & 37.56 & 29.59 & 33.34 \\
 \hline\hline
\end{tabular}
\end{table*}

\section{SUMMARY AND DISCUSSION}
\label{sec-summary}

In this work, we constructed an 
SU(3)$_L \times$ SU(3)$_R$ invariant parity doublet model based on the quark diagram.
In our model, the $(3_L, \bar{3}_R) + (\bar{3}_L, 3_R)$, $(3_L, 6_R) + (6_L, 3_R)$ and $(1_L, 8_R) + (8_L, 1_R)$ representations 
are manifestly included to describe baryon octet states from the ground states to excited states. 
We use $\chi^2$ fitting to 
determine 10 parameters in our model to  
12 physical inputs and calculate the mass spectra for hyperons with the masses smaller than 2 GeV. 
The mass spectra are well reproduced as shown in Fig.~\ref{fig-massspec}. 
Also we obtained the mixing ratio for different representations. 
For the ground-state, the results show  that,  for all the reasonable combinations of $(m^{\psi}, m^{\eta})$, the $\psi$ and $\chi$ fields are dominant, 
indicating the fact that the ground-state is well-saturated by $(3, \bar{3}) + (\bar{3}, 3) $ and $(8, 1) + (1, 8)$ representations.
This is consistent with considerations based on diquarks.

Our new model has improved the mass ordering problem in the previous analyses,
but unfortunately there arises another problem concerning with the strength of the Yukawa couplings.
We close this paper by mentioning this problem and call for further studies.

The Yukawa couplings for reasonable fit are found to be large compared to those used in our previous analyses for the two-flavor case.
The magnitude of 10 Yukawa couplings are $O(\mbox{$10$ - $30$})$ and both positive and negative signs are possible.
Our experience for two-flavor analyses indicates that
the sum and difference between the Yukawa couplings are responsible for the average and mass splitting of the positive and negative parity nucleons.
In order to explain the mass splitting of $\sim 500$ MeV, the couplings cannot be too small for whatever chiral invariant masses.
What we have not understood is why Yukawa couplings in the three-flavor case becomes larger than the two-flavor case by a few factors.

Our three-flavor model contains more representations than the two-flavor model,
and it might be possible that after diagonalization to get physical baryon spectra, 
the physical baryons have smaller Yukawa coupling,
because the Yukawa couplings for the original fields have alternating signs.
To check this possibility, we estimate the Yukawa coupling between $\sigma$ and physical nucleons obtained after diagonalization.
For this purpose we compute $\partial m_N / \partial \sigma$.

In Fig.~\ref{N_sigma}, we show the $\sigma$ dependence of nucleon mass for different generations. 
The value of $\sigma$ dependence of nucleon mass are listed in Table~\ref{tab-Yukawa}.
For each combination of $m^{\psi (\chi)}$  and $m^{\eta}$, we choose the best fitted parameters to reproduce the mass spectrum of each nucleon. 
In the upper figure, we show the $\sigma$ dependence of the first four states for two sets of $(m^{\psi (\chi)}, m^\eta)$, (600, 1000) and (800,1000) MeV. Increasing $\sigma$ from 0 to $f_\pi$, 
positive parity states show non-monotonic behavior while negative parity states grow monotonically.
This behavior is consistent with SU(2) PDM in Ref.~\cite{Marczenko:2023ohi}. Next we vary $m^\psi$ with $m^\eta$ fixed to $1000$ MeV (top left panel of Fig.~\ref{N_sigma}).
For the first four states, increasing $m^{\psi}$ reduces $\partial m_{N} / \partial \sigma$ or the Yukawa coupling to $\sigma$,
since $m^\psi$ explains the majority of $m_N$.
We also vary $m^\eta$ with $m^\psi$ fixed to $700$ MeV (top right panel of Fig.~\ref{N_sigma}), and found that the first four states are very insensitive to the details of $m_\eta$. 

For the other two highest excited states, we found the masses increase almost linearly as a function of $\sigma$.
This seems odd to us, as we expected that the highly excited states would decouple from the chiral symmetry breaking.
We guess that the overall mass scale for these states should be discussed 
in the context of other mechanisms such as stringy excitations seen in the Regge trajectories.

To summarize, our model for SU(3)$_L$ $\times$ SU(3)$_R$
has improved the description of mass ordering associated with the flavor difference,
but we found that the size of Yukawa couplings are still problematic.
It is surprising to us that imposing the SU(3)$_L$ $\times$ SU(3)$_R$ symmetry (not just SU(3)$_V$) introduces,
besides many possible baryonic representations,
several unexpected issues in constructing a Lagrangian in the linear realization of chiral symmetry.
There still seem some missing elements and further studies are called for.

\section*{Acknowledgement}

G.B. is supported by JST SPRING, Grant No. JPMJSP2125.  G.B. would like to take this opportunity to thank the
“Interdisciplinary Frontier Next-Generation Researcher Program of the Tokai
Higher Education and Research System”; 
M.H. by JSPS KAKENHI Grant No. JP20K03927 and JP23H05439; 
T.K. by JSPS KAKENHI Grant No. 23K03377 and No. 18H05407, and also by the Graduate Program on Physics for the Universe (GPPU) at Tohoku university.


\bibliography{ref_3fPDM_2022.bib}
\end{document}